\documentclass[fleqn,usenatbib,onecolumn]{mnras}

\usepackage{newtxtext,newtxmath}

\usepackage[T1]{fontenc}
\usepackage{ae,aecompl}

\usepackage{graphicx}	
\usepackage{amsmath}	
\usepackage{amssymb}	

\usepackage{bm}  

\renewcommand{\Re}{\operatorname{Re}}
\renewcommand{\Im}{\operatorname{Im}}

\title[Exact joint pseudo-$C_\ell$ likelihood]{Exact joint likelihood of pseudo-$C_\ell$ estimates from correlated Gaussian cosmological fields}

\author[R. E. Upham, L. Whittaker and M. L. Brown]{
Robin E. Upham,$^{1}$\thanks{E-mail: robin.upham@manchester.ac.uk (REU)}
Lee Whittaker$^{2, 1}$
and Michael L. Brown$^{1}$
\\
$^{1}$Jodrell Bank Centre for Astrophysics, Department of Physics and Astronomy, University of Manchester, Manchester M13 9PL, UK\\
$^{2}$Department of Physics and Astronomy, University College London, Gower Street, London WC1E 6BT, UK}

\date{
Accepted 2019 November 13. Received 2019 November 12; in original form 2019 August 02 }

\pubyear{2019}

\begin{document}
\label{firstpage}
\pagerange{\pageref{firstpage}--\pageref{lastpage}}
\maketitle

\begin{abstract}
We present the exact joint likelihood of pseudo-$C_\ell$ power spectrum estimates measured from an arbitrary number of Gaussian cosmological fields. Our method is applicable to both spin-0 fields and spin-2 fields, including a mixture of the two, and is relevant to Cosmic Microwave Background, weak lensing and galaxy clustering analyses. We show that Gaussian cosmological fields are mixed by a mask in such a way that retains their Gaussianity and derive exact expressions for the covariance of the cut-sky spherical harmonic coefficients, the pseudo-$a_{\ell m}$s, without making any assumptions about the mask geometry. We then show that each auto- or cross-pseudo-$C_\ell$ estimator can be written as a quadratic form, and apply the known joint distribution of quadratic forms to obtain the exact joint likelihood of a set of pseudo-$C_\ell$ estimates in the presence of an arbitrary mask. We show that the same formalism can be applied to obtain the exact joint likelihood of Quadratic Maximum Likelihood power spectrum estimates. Considering the polarisation of the Cosmic Microwave Background as an example, we show using simulations that our likelihood recovers the full, exact multivariate distribution of $EE$, $BB$ and $EB$ pseudo-$C_\ell$ power spectra. Our method provides a route to robust cosmological constraints from future Cosmic Microwave Background and large-scale structure surveys in an era of ever-increasing statistical precision.
\end{abstract}




\graphicspath{{Figs/}}

\section{Introduction}

In the age of precision cosmology, all possible sources of bias in the process of cosmological inference must be examined and controlled. One such source of bias that has recently been receiving an increasing amount of attention is the choice of likelihood function, which is an essential ingredient of Bayesian inference. The traditional, convenient choice is a multivariate Gaussian likelihood. Depending on the cosmological observable in question, this assumption may be anything from (near-)exact to severely wrong. In the latter case, this error can potentially propagate through to biased cosmological constraints.

Here we consider the likelihood of observed cosmological power spectra. Even in contemporary analyses these are routinely assumed to be Gaussian distributed \citep[e.g.][]{Hikage2018, Liu2018, Planck2018V}, and the choice of covariance matrix in a Gaussian likelihood is an extremely active area of research \citep[e.g.][]{Kodwani2018, Harnois-Deraps2019}. However, it is well known that the true power spectrum likelihood is strongly non-Gaussian, especially on large scales, both for Cosmic Microwave Background \citep{Sun2013} and weak lensing observations \citep{Hartlap2009, Sellentin2018, Sellentin2018a}.

For full sky observations with isotropic noise and in the absence of systematic effects, the exact likelihood of observed power spectra is known \citep{Percival2006}. However, the situation is more complex for realistic observations, and specifically in the presence of a mask. \cite{Wandelt2001} presented the exact distribution of power in a single multipole of a single spin-$0$ field in the presence of an azimuthally symmetric mask. Here we extend this to obtain the exact likelihood of an arbitrary number of multipoles of an arbitrary number of correlated spin-0 and spin-2 fields, each observed with a mask of any geometry. This is the first mathematically exact approach to such a generalisation. Previously, many approximate forms have been suggested, some of which attempt to model other realistic effects beyond a cut sky \citep{Percival2006, Hamimeche2008, Mangilli2015, Kalus2016}. Alternative approaches include Gaussianisation of the data vector \citep[e.g.][]{Wang2018} or likelihood-free inference \citep{Alsing2019, Taylor2019}.

In this paper we will consider the exact distribution of power spectrum estimates measured on a cut sky using the pseudo-$C_\ell$ estimator. This estimator has the advantage of speed as it relies only on spherical harmonic transforms, which have fast implementations in software packages such as {\tt HEALPix} \citep{Gorski2005}. It has been used in the Hyper Suprime-Cam first-year weak lensing analysis in \cite{Hikage2018} (under the flat-sky approximation using Fourier transforms in place of spherical harmonic transforms) and is also planned to be used for the weak lensing analysis of future {\it Euclid} data. With the statistical power offered by next-generation experiments such as {\it Euclid}, it is essential to understand the likelihood of all estimators in order to ensure reliable cosmological constraints are obtained. Alternatives to the pseudo-$C_\ell$ estimator include the Quadratic Maximum Likelihood (QML) estimator \citep{Tegmark1997}, which is designed to return minimum-variance estimates provided a suitable choice of fiducial model is made. In Section~\ref{Sec:pseudo_alm_dist} we derive the distribution of observed spherical harmonic coefficients, the pseudo-$a_{\ell m}$s, show that their Gaussianity is preserved in the presence of a mask---regardless of its geometry--- and present exact expressions for the covariance of any pair of pseudo-$a_{\ell m}$s. In Section \ref{Sec:pcl_dist} we show that each pseudo-$C_\ell$ estimator may be written as a quadratic form in the pseudo-$a_{\ell m}$s, and use the known joint distribution of such quadratic forms to derive the full, exact joint likelihood of pseudo-$C_\ell$ estimates. We also show that the same formalism can be applied to obtain the exact joint likelihood of QML estimates. We consider the polarisation of the Cosmic Microwave Background (CMB) as an example in Sections \ref{Sec:examples} and \ref{Sec:results}, where we demonstrate that our pseudo-$C_\ell$ likelihood reproduces the exact distribution of $EE$, $BB$ and $EB$ pseudo-$C_\ell$ power spectra in the presence of a general mask. We discuss the implications of our work and present our conclusions in Section \ref{Sec:conclusions}.

\section[Pseudo-alm distribution]{Pseudo-$\lowercase{a}_{\ell \lowercase{m}}$ distribution}
\label{Sec:pseudo_alm_dist}

In this section we derive the distribution of spherical harmonic coefficients on a general cut sky, the pseudo-$a_{\ell m}$s.

\subsection[Full-sky alm distribution]{Full-sky $a_{\ell m}$ distribution}
\label{Sec:Full_sky_alm_dist}

We begin by considering the distribution of $a_{\ell m}$s on the full sky. We assume correlated, isotropic Gaussian spin-0 or spin-2 cosmological fields $\alpha \left( \Omega \right)$, where $\Omega$ represents sky coordinates. We will use lower case Greek characters to represent cosmological fields throughout. Each spin-0 field can be decomposed in terms of spherical harmonics as
\begin{equation}
    \alpha \left( \Omega \right) =
    \sum_{\ell = 0}^\infty \sum_{m = -\ell}^\ell
    a_{\ell m}^{\left( \alpha \right)} Y_{\ell m} \left( \Omega \right),
\end{equation}
where $a_{\ell m}^{\left( \alpha \right)}$ are the spherical harmonic coefficients for field $\alpha$. Each (complex) spin-2 field can be decomposed in terms of spin-weighted spherical harmonics as
\begin{equation}
    \left.\begin{aligned}
    \alpha \left( \Omega \right) &=
    \sum_{\ell = 0}^\infty \sum_{m = -\ell}^\ell
    \left( a_{\ell m}^{E \left( \alpha \right)} + i a_{\ell m}^{B \left( \alpha \right)} \right)
    {}_2Y_{\ell m} \left( \Omega \right), \\
    \alpha^* \left( \Omega \right) &=
    \sum_{\ell = 0}^\infty \sum_{m = -\ell}^\ell
    \left( a_{\ell m}^{E \left( \alpha \right)} - i a_{\ell m}^{B \left( \alpha \right)} \right)
    {}_{-2}Y_{\ell m} \left( \Omega \right),
    \end{aligned}\right.
\end{equation}
where the superscript $*$ denotes complex conjugation, and $a_{\ell m}^{E \left( \alpha \right)}$ and $a_{\ell m}^{B \left( \alpha \right)}$ are the spherical harmonic coefficients of the $E$- and $B$-mode components of field $\alpha$, respectively.

For spin-0 and spin-2 spherical harmonics, the $a_{\ell m}$s have the property that
\begin{equation}
    a_{\ell -m}^{ \left( \alpha \right) } =
    \left( -1 \right)^m \left( a_{\ell m}^{ \left( \alpha \right) } \right)^*,
    \label{Eqn:alm_symmetry}
\end{equation}
which further implies that the $m = 0$ $a_{\ell m}$s must be real. For a given multipole $\ell$, the real and imaginary parts of $a_{\ell m}^{\left( \alpha \right)}$ for all $m > 0$ are  independently and identically distributed as multivariate Gaussian with mean $\mathbfit{0}$ and covariance matrix $\frac{1}{2} \mathbfss{C}_\ell$, where the elements of $\mathbfss{C}_\ell$ are given by
\begin{equation}
    \mathbfss{C}_\ell^{\alpha \beta} = C_\ell^{\alpha \beta}.
\end{equation}
Here, each of $\alpha$ and $\beta$ can represent either a spin-0 field or the $E$- or $B$-mode component of a spin-2 field, and $C_\ell^{\alpha \beta}$ is the underlying angular cross-power spectrum between fields $\alpha$ and $\beta$ (which may be 0). $\mathbfss{C}_\ell$ is symmetric such that $C_\ell^{\alpha \beta} = C_\ell^{\beta \alpha}$, and $\alpha = \beta$ gives the auto-power spectrum $C_\ell^{\alpha \alpha}$. In the case of $m = 0$, the lack of an imaginary part means that all of the underlying power is in the real part, and hence the $a_{\ell 0}^{\left( \alpha \right)}$ are multivariate Gaussian distributed with mean $\mathbfit{0}$ and covariance matrix $\mathbfss{C}_\ell$. The real part of each $a_{\ell m}^{\left( \alpha \right)}$ is independent from all imaginary parts, and vice versa. All $a_{\ell m}^{\left( \alpha \right)}$ are independent between different $\ell$ and $m$; the only cross-correlation is between $a_{\ell m}^{\left( \alpha \right)}$s having the same $\ell$ and the same $m$, but different fields $\alpha$. The $m < 0$ $a_{\ell m}$s can be regarded as deterministic functions of their positive-$m$ counterparts following Equation \eqref{Eqn:alm_symmetry}, rather than separate random variables.

\subsection{Effect of a cut sky}

In real space, the effect of a mask is to multiply each field by a window function $W_\alpha \left( \Omega \right)$, which can in general be unique to each field:
\begin{equation}
    \widetilde{\alpha} \left( \Omega \right) = W_\alpha \left( \Omega \right) \alpha \left( \Omega \right).
\end{equation}
The multiplication in real space is equivalent to a convolution in spherical harmonic space, which has the effect of mixing the $a_{\ell m}$s between different $\ell$, $m$, and between $E$- and $B$-modes in the case of spin-2 fields, to give the pseudo-$a_{\ell m}$s \citep{Lewis2001, Brown2005}:
\begin{align}
    &\widetilde{a}_{\ell m}^{\left( \alpha \right)} =
    \sum_{\ell' = 0}^\infty \sum_{m' = -\ell'}^{\ell'}
    {}_0W_{\ell \ell'}^{m m'} a^{\left( \alpha \right)}_{\ell' m'};
    \label{Eqn:pseudo_alm_spin0}
    \\
    &\widetilde{a}_{\ell m}^{E \left( \alpha \right)} =
    \sum_{\ell' = 0}^\infty \sum_{m' = -\ell'}^{\ell'} \left(
    W^+_{\ell \ell' m m'} a^{E \left( \alpha \right)}_{\ell' m'}
    + W^-_{\ell \ell' m m'} a^{B \left( \alpha \right)}_{\ell' m'}
    \right);
    \label{Eqn:pseudo_alm_emode}
    \\
    &\widetilde{a}_{\ell m}^{B \left( \alpha \right)} =
    \sum_{\ell' = 0}^\infty \sum_{m' = -\ell'}^{\ell'} \left(
    W^+_{\ell \ell' m m'} a^{B \left( \alpha \right)}_{\ell' m'}
    - W^-_{\ell \ell' m m'} a^{E \left( \alpha \right)}_{\ell' m'}
    \right).
    \label{Eqn:pseudo_alm_bmode}
\end{align}
The spin-weighted spherical harmonic space window functions are given by
\begin{equation}
    _sW_{\ell \ell'}^{m m'} = \int d\Omega
    _sY_{\ell' m'} \left( \Omega \right) W_\alpha \left( \Omega \right) _sY_{\ell m}^* \left( \Omega \right),
    \label{Eqn:harm_window}
\end{equation}
where we note that the optional field-dependence of the mask reflected in $W_\alpha \left( \Omega \right)$ also means that each $_sW_{\ell \ell'}^{m m'}$ is implicitly field-specific, but we drop the $\alpha$ on the left-hand side to limit the number of indices. Following \citet{Brown2005} we have defined
\begin{equation}
    W_{\ell \ell' m m'}^+ =
    \frac{1}{2} \left( _2W_{\ell \ell'}^{m m'} + _{-2}W_{\ell \ell'}^{m m'} \right); \quad \quad
    W_{\ell \ell' m m'}^- =
    \frac{i}{2} \left( _2W_{\ell \ell'}^{m m'} - _{-2}W_{\ell \ell'}^{m m'} \right),
\end{equation}
which may also be specific to each field.

\subsection[Pseudo-alm distribution]{Pseudo-$a_{\ell m}$ distribution}

Equations~\eqref{Eqn:pseudo_alm_spin0}--\eqref{Eqn:pseudo_alm_bmode} describe the effects of a mask on spin-0 and spin-2 fields. More generally, any observed pseudo-$a_{\ell m}$ can be written as a linear combination of full-sky $a_{\ell m}$s as
\begin{equation}
    \widetilde{a}_{\ell m}^{\left( \alpha \right)} = \sum_{\ell' m'} \sum_\beta
    \frac{\partial \widetilde{a}_{\ell m}^{\left( \alpha \right)}}
    {\partial a_{\ell' m'}^{\left( \beta \right)}}
    a_{\ell' m'}^{\left( \beta \right)},
\label{Eqn:pseudo_alm_as_sum}
\end{equation}
where $\alpha$ and $\beta$ may each be either a single spin-0 field or the $E$- or $B$-mode component of a spin-2 field. We can expand this into real and imaginary parts as $\widetilde{a}_{\ell m}^{\left( \alpha \right)} = \Re \left( \widetilde{a}_{\ell m}^{\left( \alpha \right)} \right) + i \Im \left( \widetilde{a}_{\ell m}^{\left( \alpha \right)} \right)$, and we show in Appendix \ref{App:cov_derivation} that the respective parts are given by
\begin{equation}
\begin{split}
\Re \left( \widetilde{a}_{\ell m}^{ \left( \alpha \right) } \right) =
\sum_{\beta, \ell'} \Bigg[ &
\Re \left(
\frac{\partial \widetilde{a}_{\ell m}^{ \left( \alpha \right) }}
{\partial a_{\ell' 0}^{ \left( \beta \right) }}
\right)
\Re \left( a_{\ell' 0}^{ \left( \beta \right) } \right) \\
& +
\sum_{m' > 0} \Bigg( \left[
\Re \left(
\frac{\partial \widetilde{a}_{\ell m}^{ \left( \alpha \right) }}
{\partial a_{\ell' m'}^{ \left( \beta \right) }}
\right)
+ \left( -1 \right)^{m'}
\Re \left(
\frac{\partial \widetilde{a}_{\ell m}^{ \left( \alpha \right) }}
{\partial a_{\ell' -m'}^{ \left( \beta \right) }}
\right) \right]
\Re \left(
a_{\ell' m'}^{ \left( \beta \right) }
\right)
- \left[ \Im \left(
\frac{\partial \widetilde{a}_{\ell m}^{ \left( \alpha \right) }}
{\partial a_{\ell' m'}^{ \left( \beta \right) }}
\right)
- \left( -1 \right)^{m'}
\Im \left(
\frac{\partial \widetilde{a}_{\ell m}^{ \left( \alpha \right) }}
{\partial a_{\ell' -m'}^{ \left( \beta \right) }}
\right) \right]
\Im \left(
a_{\ell' m'}^{ \left( \beta \right) }
\right) \Bigg) \Bigg];
\label{Eqn:Re_alm_general}
\end{split}
\end{equation}
\begin{equation}
\begin{split}
\Im \left( \widetilde{a}_{\ell m}^{ \left( \alpha \right) } \right) =
\sum_{\beta, \ell'} \Bigg[ &
\Im \left(
\frac{\partial \widetilde{a}_{\ell m}^{ \left( \alpha \right) }}
{\partial a_{\ell' 0}^{ \left( \beta \right) }}
\right)
\Re \left( a_{\ell' 0}^{ \left( \beta \right) } \right) \\
& +
\sum_{m' > 0} \Bigg( \left[
\Im \left(
\frac{\partial \widetilde{a}_{\ell m}^{ \left( \alpha \right) }}
{\partial a_{\ell' m'}^{ \left( \beta \right) }}
\right)
+ \left( -1 \right)^{m'}
\Im \left(
\frac{\partial \widetilde{a}_{\ell m}^{ \left( \alpha \right) }}
{\partial a_{\ell' -m'}^{ \left( \beta \right) }}
\right) \right]
\Re \left(
a_{\ell' m'}^{ \left( \beta \right) }
\right) +
\left[ \Re \left(
\frac{\partial \widetilde{a}_{\ell m}^{ \left( \alpha \right) }}
{\partial a_{\ell' m'}^{ \left( \beta \right) }}
\right)
- \left( -1 \right)^{m'}
\Re \left(
\frac{\partial \widetilde{a}_{\ell m}^{ \left( \alpha \right) }}
{\partial a_{\ell' -m'}^{ \left( \beta \right) }}
\right) \right]
\Im \left(
a_{\ell' m'}^{ \left( \beta \right) }
\right) \Bigg) \Bigg].
\label{Eqn:Im_alm_general}
\end{split}
\end{equation}
Each derivative in these equations is a constant weight that is uniquely specified by the geometry of the mask, and is independent of each $a_{\ell' m'}^{\left( \beta \right)}$. Therefore, Equations~\eqref{Eqn:Re_alm_general} and \eqref{Eqn:Im_alm_general} are simply linear combinations of Gaussian random variables. Any linear combination of Gaussian variables is itself Gaussian distributed, with mean and variance adding linearly such that
\begin{equation}
    Y = \sum_i a_i X_i, \quad X_i \sim \mathcal{N} \left( \mu_i, \sigma_i^2 \right)
    \quad \implies \quad
    Y \sim \mathcal{N} \left( \sum_i a_i \mu_i,
    \sum_{i i'} a_i a_{i'} ~ \mathrm{Cov} \left( X_i, X_{i'} \right) \right),
    \label{Eqn:sum_of_gaussians}
\end{equation}
where we have followed the convention that $\sim$ means ``distributed as'' and $\mathcal{N} \left( \mu, \sigma^2 \right)$ represents a univariate Gaussian distribution with mean $\mu$ and variance $\sigma^2$. $\mathrm{Cov} \left( \cdot, \cdot \right)$ represents the covariance of two random variables, and the covariance of any variable with itself is its variance. From Equation \eqref{Eqn:sum_of_gaussians} we find that the real and imaginary parts of $\widetilde{a}_{\ell m}^{\left( \alpha \right)}$ each follow a Gaussian distribution with zero mean, since the full-sky $a_{\ell m}$s each have zero mean. We show in Appendix \ref{App:cov_derivation} that the variance of the real and imaginary part of each pseudo-$a_{\ell m}$ is
\begin{equation}
\begin{split}
\mathrm{Var} \left(
\Re \left( \widetilde{a}_{\ell m}^{ \left( \alpha \right) } \right)
\right) =
\sum_{\beta, \gamma}
\sum_{\ell'}
C_{\ell'}^{\beta \gamma}
\Bigg[
\Re \left(
\frac{\partial \widetilde{a}_{\ell m}^{ \left( \alpha \right) }}
{\partial a_{\ell' 0}^{ \left( \beta \right) }}
\right)
\Re \left(
\frac{\partial \widetilde{a}_{\ell m}^{ \left( \alpha \right) }}
{\partial a_{\ell' 0}^{ \left( \gamma \right) }}
\right)
+ \frac{1}{2}
\sum_{m' > 0} \Bigg( &
\Re \left(
\left[
\frac{\partial \widetilde{a}_{\ell m}^{ \left( \alpha \right) }}
{\partial a_{\ell' m'}^{ \left( \beta \right) }}
\right]^*
\frac{\partial \widetilde{a}_{\ell m}^{ \left( \alpha \right) }}
{\partial a_{\ell' m'}^{ \left( \gamma \right) }}
\right)
+ \Re \left(
\left[
\frac{\partial \widetilde{a}_{\ell m}^{ \left( \alpha \right) }}
{\partial a_{\ell' -m'}^{ \left( \beta \right) }}
\right]^*
\frac{\partial \widetilde{a}_{\ell m}^{ \left( \alpha \right) }}
{\partial a_{\ell' -m'}^{ \left( \gamma \right) }}
\right) \\
&+
\left( -1 \right)^{m'} \left[
\Re \left(
\frac{\partial \widetilde{a}_{\ell m}^{ \left( \alpha \right) }}
{\partial a_{\ell' m'}^{ \left( \beta \right) }}
\frac{\partial \widetilde{a}_{\ell m}^{ \left( \alpha \right) }}
{\partial a_{\ell' -m'}^{ \left( \gamma \right) }}
\right)
+ \Re \left(
\frac{\partial \widetilde{a}_{\ell m}^{ \left( \alpha \right) }}
{\partial a_{\ell' -m'}^{ \left( \beta \right) }}
\frac{\partial \widetilde{a}_{\ell m}^{ \left( \alpha \right) }}
{\partial a_{\ell' m'}^{ \left( \gamma \right) }}
\right) \right]
\Bigg) \Bigg];
\label{Eqn:re_alm_variance}
\end{split}
\end{equation}
\begin{equation}
\begin{split}
\mathrm{Var} \left(
\Im \left( \widetilde{a}_{\ell m}^{ \left( \alpha \right) } \right)
\right) =
\sum_{\beta, \gamma}
\sum_{\ell'}
C_{\ell'}^{\beta \gamma}
\Bigg[
\Im \left(
\frac{\partial \widetilde{a}_{\ell m}^{ \left( \alpha \right) }}
{\partial a_{\ell' 0}^{ \left( \beta \right) }}
\right)
\Im \left(
\frac{\partial \widetilde{a}_{\ell m}^{ \left( \alpha \right) }}
{\partial a_{\ell' 0}^{ \left( \gamma \right) }}
\right)
+ \frac{1}{2}
\sum_{m' > 0} \Bigg( &
\Re \left(
\left[
\frac{\partial \widetilde{a}_{\ell m}^{ \left( \alpha \right) }}
{\partial a_{\ell' m'}^{ \left( \beta \right) }}
\right]^*
\frac{\partial \widetilde{a}_{\ell m}^{ \left( \alpha \right) }}
{\partial a_{\ell' m'}^{ \left( \gamma \right) }}
\right)
+ \Re \left(
\left[
\frac{\partial \widetilde{a}_{\ell m}^{ \left( \alpha \right) }}
{\partial a_{\ell' -m'}^{ \left( \beta \right) }}
\right]^*
\frac{\partial \widetilde{a}_{\ell m}^{ \left( \alpha \right) }}
{\partial a_{\ell' -m'}^{ \left( \gamma \right) }}
\right) \\
&-
\left( -1 \right)^{m'} \left[
\Re \left(
\frac{\partial \widetilde{a}_{\ell m}^{ \left( \alpha \right) }}
{\partial a_{\ell' m'}^{ \left( \beta \right) }}
\frac{\partial \widetilde{a}_{\ell m}^{ \left( \alpha \right) }}
{\partial a_{\ell' -m'}^{ \left( \gamma \right) }}
\right)
+ \Re \left(
\frac{\partial \widetilde{a}_{\ell m}^{ \left( \alpha \right) }}
{\partial a_{\ell' -m'}^{ \left( \beta \right) }}
\frac{\partial \widetilde{a}_{\ell m}^{ \left( \alpha \right) }}
{\partial a_{\ell' m'}^{ \left( \gamma \right) }}
\right) \right]
\Bigg) \Bigg],
\label{Eqn:im_alm_variance}
\end{split}
\end{equation}
where $\beta$ and $\gamma$ are all fields correlated with field $\alpha$, each of which may be either spin-0 or the $E$- or $B$-mode component of a spin-2 field.

Since both the real and imaginary parts of each pseudo-$a_{\ell m}$ are Gaussian distributed, the joint distribution for both the real and imaginary parts of all pseudo-$a_{\ell m}$s for all fields, contained in the vector $\widetilde{\mathbfit{a}}$, can be described by a multivariate Gaussian distribution,
\begin{equation}
    \widetilde{\mathbfit{a}} \sim
    \mathcal{N} \left( \mathbfit{0}, \bm{\Sigma} \right),
\end{equation}
with covariance matrix $\bm{\Sigma}$ whose diagonal elements are given by Equations \eqref{Eqn:re_alm_variance} and \eqref{Eqn:im_alm_variance}. The off-diagonal elements can be calculated using the rule---which is not specific to the Gaussian distribution---that the covariance of two linear combinations of random variables is given by
\begin{equation}
    \mathrm{Cov} \left( \sum_i \alpha_i X_i, \sum_j \beta_j Y_j \right) =
    \sum_{ij} \alpha_i \beta_j \mathrm{Cov} \left( X_i, Y_j \right),
    \label{Eqn:cov_addition}
\end{equation}
where $\alpha_i$ and $\beta_j$ are constant weights. We show in Appendix \ref{App:cov_derivation} that this leads to the following expressions for the elements of $\bm{\Sigma}$:
\begin{equation}
\begin{split}
\mathrm{Cov} \left(
\Re \left( \widetilde{a}_{\ell m}^{ \left( \alpha \right) } \right),
\Re \left( \widetilde{a}_{\ell' m'}^{ \left( \beta \right) } \right)
\right) =
\sum_{\gamma, \varepsilon}
\sum_{\ell''}
C_{\ell''}^{\gamma \varepsilon}
\Bigg[ &
\Re \left(
\frac{\partial \widetilde{a}_{\ell m}^{ \left( \alpha \right) }}
{\partial a_{\ell'' 0}^{ \left( \gamma \right) }}
\right)
\Re \left(
\frac{\partial \widetilde{a}_{\ell' m'}^{ \left( \beta \right) }}
{\partial a_{\ell'' 0}^{ \left( \varepsilon \right) }}
\right) \\
&+
\frac{1}{2}
\sum_{m'' > 0} \Bigg(
\Re \left(
\left[
\frac{\partial \widetilde{a}_{\ell m}^{ \left( \alpha \right) }}
{\partial a_{\ell'' m''}^{ \left( \gamma \right) }}
\right]^*
\frac{\partial \widetilde{a}_{\ell' m'}^{ \left( \beta \right) }}
{\partial a_{\ell'' m''}^{ \left( \varepsilon \right) }}
\right)
+ \Re \left(
\left[
\frac{\partial \widetilde{a}_{\ell m}^{ \left( \alpha \right) }}
{\partial a_{\ell'' -m''}^{ \left( \gamma \right) }}
\right]^*
\frac{\partial \widetilde{a}_{\ell' m'}^{ \left( \beta \right) }}
{\partial a_{\ell'' -m''}^{ \left( \varepsilon \right) }}
\right) \\
&~~~~~~~~~~~~~~~~~~~ +
\left( -1 \right)^{m''} \left[
\Re \left(
\frac{\partial \widetilde{a}_{\ell m}^{ \left( \alpha \right) }}
{\partial a_{\ell'' m''}^{ \left( \gamma \right) }}
\frac{\partial \widetilde{a}_{\ell' m'}^{ \left( \beta \right) }}
{\partial a_{\ell'' -m''}^{ \left( \varepsilon \right) }}
\right)
+ \Re \left(
\frac{\partial \widetilde{a}_{\ell m}^{ \left( \alpha \right) }}
{\partial a_{\ell'' -m''}^{ \left( \gamma \right) }}
\frac{\partial \widetilde{a}_{\ell' m'}^{ \left( \beta \right) }}
{\partial a_{\ell'' m''}^{ \left( \varepsilon \right) }}
\right) \right]
\Bigg) \Bigg];
\label{Eqn:cov_re_re_general}
\end{split}
\end{equation}
\begin{equation}
\begin{split}
\mathrm{Cov} \left(
\Im \left( \widetilde{a}_{\ell m}^{ \left( \alpha \right) } \right),
\Im \left( \widetilde{a}_{\ell' m'}^{ \left( \beta \right) } \right)
\right) =
\sum_{\gamma, \varepsilon}
\sum_{\ell''}
C_{\ell''}^{\gamma \varepsilon}
\Bigg[ &
\Im \left(
\frac{\partial \widetilde{a}_{\ell m}^{ \left( \alpha \right) }}
{\partial a_{\ell'' 0}^{ \left( \gamma \right) }}
\right)
\Im \left(
\frac{\partial \widetilde{a}_{\ell' m'}^{ \left( \beta \right) }}
{\partial a_{\ell'' 0}^{ \left( \varepsilon \right) }}
\right) \\
&+
\frac{1}{2}
\sum_{m'' > 0} \Bigg(
\Re \left(
\left[
\frac{\partial \widetilde{a}_{\ell m}^{ \left( \alpha \right) }}
{\partial a_{\ell'' m''}^{ \left( \gamma \right) }}
\right]^*
\frac{\partial \widetilde{a}_{\ell' m'}^{ \left( \beta \right) }}
{\partial a_{\ell'' m''}^{ \left( \varepsilon \right) }}
\right)
+ \Re \left(
\left[
\frac{\partial \widetilde{a}_{\ell m}^{ \left( \alpha \right) }}
{\partial a_{\ell'' -m''}^{ \left( \gamma \right) }}
\right]^*
\frac{\partial \widetilde{a}_{\ell' m'}^{ \left( \beta \right) }}
{\partial a_{\ell'' -m''}^{ \left( \varepsilon \right) }}
\right) \\
&~~~~~~~~~~~~~~~~~~~ -
\left( -1 \right)^{m''} \left[
\Re \left(
\frac{\partial \widetilde{a}_{\ell m}^{ \left( \alpha \right) }}
{\partial a_{\ell'' m''}^{ \left( \gamma \right) }}
\frac{\partial \widetilde{a}_{\ell' m'}^{ \left( \beta \right) }}
{\partial a_{\ell'' -m''}^{ \left( \varepsilon \right) }}
\right)
+ \Re \left(
\frac{\partial \widetilde{a}_{\ell m}^{ \left( \alpha \right) }}
{\partial a_{\ell'' -m''}^{ \left( \gamma \right) }}
\frac{\partial \widetilde{a}_{\ell' m'}^{ \left( \beta \right) }}
{\partial a_{\ell'' m''}^{ \left( \varepsilon \right) }}
\right) \right]
\Bigg) \Bigg];
\label{Eqn:cov_im_im_general}
\end{split}
\end{equation}
\begin{equation}
\begin{split}
\mathrm{Cov} \left(
\Re \left( \widetilde{a}_{\ell m}^{ \left( \alpha \right) } \right),
\Im \left( \widetilde{a}_{\ell' m'}^{ \left( \beta \right) } \right)
\right) =
\sum_{\gamma, \varepsilon}
\sum_{\ell''}
C_{\ell''}^{\gamma \varepsilon}
\Bigg[ &
\Re \left(
\frac{\partial \widetilde{a}_{\ell m}^{ \left( \alpha \right) }}
{\partial a_{\ell'' 0}^{ \left( \gamma \right) }}
\right)
\Im \left(
\frac{\partial \widetilde{a}_{\ell' m'}^{ \left( \beta \right) }}
{\partial a_{\ell'' 0}^{ \left( \varepsilon \right) }}
\right) \\
&+
\frac{1}{2}
\sum_{m'' > 0} \Bigg(
\Im \left(
\left[
\frac{\partial \widetilde{a}_{\ell m}^{ \left( \alpha \right) }}
{\partial a_{\ell'' m''}^{ \left( \gamma \right) }}
\right]^*
\frac{\partial \widetilde{a}_{\ell' m'}^{ \left( \beta \right) }}
{\partial a_{\ell'' m''}^{ \left( \varepsilon \right) }}
\right)
+ \Im \left(
\left[
\frac{\partial \widetilde{a}_{\ell m}^{ \left( \alpha \right) }}
{\partial a_{\ell'' -m''}^{ \left( \gamma \right) }}
\right]^*
\frac{\partial \widetilde{a}_{\ell' m'}^{ \left( \beta \right) }}
{\partial a_{\ell'' -m''}^{ \left( \varepsilon \right) }}
\right) \\
&~~~~~~~~~~~~~~~~~~~ +
\left( -1 \right)^{m''} \left[
\Im \left(
\frac{\partial \widetilde{a}_{\ell m}^{ \left( \alpha \right) }}
{\partial a_{\ell'' m''}^{ \left( \gamma \right) }}
\frac{\partial \widetilde{a}_{\ell' m'}^{ \left( \beta \right) }}
{\partial a_{\ell'' -m''}^{ \left( \varepsilon \right) }}
\right)
+ \Im \left(
\frac{\partial \widetilde{a}_{\ell m}^{ \left( \alpha \right) }}
{\partial a_{\ell'' -m''}^{ \left( \gamma \right) }}
\frac{\partial \widetilde{a}_{\ell' m'}^{ \left( \beta \right) }}
{\partial a_{\ell'' m''}^{ \left( \varepsilon \right) }}
\right) \right]
\Bigg) \Bigg].
\label{Eqn:cov_re_im_general}
\end{split}
\end{equation}

This is our first key result: that the spherical harmonic coefficients of correlated Gaussian fields measured on a cut sky, which follow a multivariate Gaussian distribution, have mean $\mathbfit{0}$ and covariance matrix $\bm{\Sigma}$ whose elements are given by Equations \eqref{Eqn:cov_re_re_general}--\eqref{Eqn:cov_re_im_general}. For a given cosmological model and survey mask, these expressions can then be applied to a particular set of observational probes by specifying the fields in question ($\alpha, \beta$), the cosmological signals ($C_\ell^{\gamma\epsilon}$), and all non-zero derivative terms, which once again are completely specified by the geometry of the survey mask. (We demonstrate this explicitly for the case of the CMB later in Section~\ref{Sec:CMB_alm_covariance}.)

\section[Pseudo-Cl distribution]{Pseudo-$C_\ell$ distribution}
\label{Sec:pcl_dist}

In this section we derive the distribution of pseudo-$C_\ell$ estimates for auto- and cross-power spectra of an arbitrary number of correlated spin-0 and spin-2 fields, for a general cut sky.

\subsection[The pseudo-Cl estimator]{The pseudo-$C_\ell$ estimator}

On the full sky, the covariance between either real parts or imaginary parts of two $a_{\ell m}$s having the same $(\ell, m)$ is equal to the corresponding underlying power spectrum, modulo a factor of $\frac{1}{2}$ for $m > 0$ (see Section \ref{Sec:Full_sky_alm_dist}). Therefore, an unbiased estimator of the power spectrum is given by an appropriately weighted sample covariance of observed $a_{\ell m}$s:
\begin{equation}
    \widehat{C}_\ell^{\alpha \beta} = \frac{1}{2 \ell + 1} \left[
    \Re \left( a_{\ell 0}^{\left( \alpha \right)} \right)
    \Re \left( a_{\ell 0}^{\left( \beta \right)} \right)
    + 2 \sum_{m = 1}^{\ell} \left[
    \Re \left( a_{\ell m}^{\left( \alpha \right)} \right)
    \Re \left( a_{\ell m}^{\left( \beta \right)} \right)
    + \Im \left( a_{\ell m}^{\left( \alpha \right)} \right)
    \Im \left( a_{\ell m}^{\left( \beta \right)} \right)
    \right] \right]
    = \frac{1}{2 \ell + 1} \sum_{m = -\ell}^{\ell}
    a_{\ell m}^{\left( \alpha \right)}
    \left( a_{\ell m}^{\left( \beta \right)} \right)^*,
\end{equation}
where we implicitly take the real part of the result. The auto-spectrum estimator is given by the special case where $\alpha = \beta$. Under a cut sky, this becomes the pseudo-$C_\ell$ estimator \citep{Wandelt2001, Hivon2002, Brown2005}, written in terms of the pseudo-$a_{\ell m}$s:
\begin{equation}
    \widetilde{C}_\ell^{\alpha \beta} = \frac{1}{2 \ell + 1} \left[
    \Re \left( \widetilde{a}_{\ell 0}^{\left( \alpha \right)} \right)
    \Re \left( \widetilde{a}_{\ell 0}^{\left( \beta \right)} \right)
    + 2 \sum_{m = 1}^{\ell} \left[
    \Re \left( \widetilde{a}_{\ell m}^{\left( \alpha \right)} \right)
    \Re \left( \widetilde{a}_{\ell m}^{\left( \beta \right)} \right)
    + \Im \left( \widetilde{a}_{\ell m}^{\left( \alpha \right)} \right)
    \Im \left( \widetilde{a}_{\ell m}^{\left( \beta \right)} \right)
    \right] \right].
    \label{Eqn:pcl_estimator_full}
\end{equation}
We can write this as a matrix equation involving the vector of all pseudo-$a_{\ell m}$s, $\widetilde{\mathbfit{a}}$:
\begin{equation}
    \widetilde{C}_\ell^{\alpha \beta} =
    \widetilde{\mathbfit{a}}^T \mathbfss{M}_\ell^{\alpha \beta}
    \widetilde{\mathbfit{a}},
    \label{Eqn:estimator_as_qf}
\end{equation}
where $\mathbfss{M}_\ell^{\alpha \beta}$ is a real symmetric matrix chosen to pick out the correct elements of $\widetilde{\mathbfit{a}}$ to match the expression in Equation \eqref{Eqn:pcl_estimator_full}. This may not be obvious, so to demonstrate we will choose an order for $\widetilde{\mathbfit{a}}$ and present the corresponding form for $\mathbfss{M}_\ell^{\alpha \beta}$, but we point out that this order is arbitrary and could be chosen, for example, to optimise computation of the likelihood. For $N$ correlated fields, each of which is either a spin-0 field or the $E$- or $B$-mode component of a spin-2 field, and a maximum measured multipole of $\ell_\mathrm{max}$, we can decompose $\widetilde{\mathbfit{a}}$ into three hierarchical levels as
\begin{equation}
    \widetilde{\mathbfit{a}} =
    \begin{pmatrix}
        \widetilde{\mathbfit{a}}^{\left( \mathrm{field~1} \right)} \\
        \widetilde{\mathbfit{a}}^{\left( \mathrm{field~2} \right)} \\
        \vdots \\
        \widetilde{\mathbfit{a}}^{\left( \mathrm{field~}N \right)} \\
    \end{pmatrix};
    \quad
    \widetilde{\mathbfit{a}}^{\left( \mathrm{field~}\alpha \right)} =
    \begin{pmatrix}
        \widetilde{\mathbfit{a}}_{\ell = 0}^{\left( \alpha \right)} \\
        \widetilde{\mathbfit{a}}_{\ell = 1}^{\left( \alpha \right)} \\
        \vdots \\
        \widetilde{\mathbfit{a}}_{\ell_\mathrm{max}}^{\left( \alpha \right)} \\
    \end{pmatrix};
    \quad
    \widetilde{\mathbfit{a}}_\ell^{\left( \alpha \right)} =
    \begin{pmatrix}
        \Re \left( \widetilde{a}_{\ell 0}^{\left( \alpha \right)} \right) \\
        \Re \left( \widetilde{a}_{\ell 1}^{\left( \alpha \right)} \right) \\
        \Im \left( \widetilde{a}_{\ell 1}^{\left( \alpha \right)} \right) \\
        \Re \left( \widetilde{a}_{\ell 2}^{\left( \alpha \right)} \right) \\
        \Im \left( \widetilde{a}_{\ell 2}^{\left( \alpha \right)} \right) \\
        \vdots \\
        \Re \left( \widetilde{a}_{\ell \ell}^{\left( \alpha \right)} \right) \\
        \Im \left( \widetilde{a}_{\ell \ell}^{\left( \alpha \right)} \right) \\
    \end{pmatrix}.
    \label{Eqn:a_hierarchy}
\end{equation}
We can then similarly decompose $\mathbfss{M}_\ell^{\alpha \beta}$ into blocks to pick out the correct elements:
\begin{align}
&\mathbfss{M}_\ell^{\alpha \beta} =
\begin{array}{r l} \mathit{Field:} &
    \begin{array}{c c c c c c c c}
    \hphantom{~\,}1 & 2 & \ldots & \hphantom{~\,}\alpha & \hphantom{\,}\ldots & \hphantom{~\,}\beta & \hphantom{\,}\ldots & N
    \end{array} \\
    \begin{array}{r}
    1 \\ 2 \\ \vdots \\ \alpha \\ \vdots \\ \beta \\ \vdots \\ N
    \end{array} &
    \left(
    \begin{array}{c c c c c c c c}
    \mathbfss{0} & \mathbfss{0} & \ldots & \mathbfss{0} & \ldots & \mathbfss{0} & \ldots & \mathbfss{0} \\
    \mathbfss{0} & \mathbfss{0} & \ldots & \mathbfss{0} & \ldots & \mathbfss{0} & \ldots & \mathbfss{0} \\
    \vdots & \vdots & \ddots & \vdots & \ddots & \vdots & \ddots & \vdots \\
    \mathbfss{0} & \mathbfss{0} & \ldots & \mathbfss{0} & \ldots & \mathbfss{M}_\ell & \ldots & \mathbfss{0} \\
    \vdots & \vdots & \ddots & \vdots & \ddots & \vdots & \ddots & \vdots \\
    \mathbfss{0} & \mathbfss{0} & \ldots & \mathbfss{M}_\ell & \ldots & \mathbfss{0} & \ldots & \mathbfss{0} \\
    \vdots & \vdots & \ddots & \vdots & \ddots & \vdots & \ddots & \vdots \\
    \mathbfss{0} & \mathbfss{0} & \ldots & \mathbfss{0} & \ldots & \mathbfss{0} & \ldots & \mathbfss{0}
    \end{array}
    \right)
\end{array};
\quad
\mathbfss{M}_\ell =
\begin{array}{c l} \ell': &
    \begin{array}{c c c c c c c}
    \hphantom{~\,}0 & 1 & \ldots & \hphantom{~}\ell & \hphantom{\,}\ldots & \hspace{-0.5em}\ell_\mathrm{max}
    \end{array} \\
    \begin{array}{c}
    0 \\ 1 \\ \vdots \\ \ell \\ \vdots \\ \ell_\mathrm{max}
    \end{array} &
    \left(
    \begin{array}{c c c c c c c}
    \mathbfss{0} & \mathbfss{0} & \ldots & \mathbfss{0} & \ldots & \mathbfss{0} \\
    \mathbfss{0} & \mathbfss{0} & \ldots & \mathbfss{0} & \ldots & \mathbfss{0} \\
    \vdots & \vdots & \ddots & \vdots & \ddots & \vdots \\
    \mathbfss{0} & \mathbfss{0} & \ldots & \mathbfss{M} & \ldots & \mathbfss{0} \\
    \vdots & \vdots & \ddots & \vdots & \ddots & \vdots \\
    \mathbfss{0} & \mathbfss{0} & \ldots & \mathbfss{0} & \ldots & \mathbfss{0} \\
    \end{array}
    \right)
\end{array}; \\
&\mathbfss{M} =
\left( \frac{1 + \delta_{\alpha \beta}}{2 \ell + 1}  \right)
\mathrm{diag} \left( \frac{1}{2}, 1, 1, 1, \ldots \right).
\label{Eqn:m_diag}
\end{align}
The Kronecker delta $\delta_{\alpha \beta}$ accounts for the auto-spectrum case where $\alpha = \beta$. \mathbfss{M} has $2 \ell + 1$ diagonal elements, so the size of \mathbfss{M} is dependent on $\ell$ but its form is not. Each block in $\mathbfss{M}_\ell$ has $2 \ell' + 1$ elements in each dimension, so $\mathbfss{M}_\ell$ has a total of $\sum_{\ell' = 0}^{\ell_\mathrm{max}} \left( 2 \ell' + 1 \right) = \left( \ell_\mathrm{max} + 1 \right)^2$ elements along each side. This is also true of every zero block in $\mathbfss{M}_\ell^{\alpha \beta}$, so $\mathbfss{M}_\ell^{\alpha \beta}$ has a total of $N \times \left( \ell_\mathrm{max} + 1 \right)^2$ elements along each side. We note that the choice of $\ell_\mathrm{\min} = 0$ is arbitrary, and any other $\ell_\mathrm{\min}$ could be chosen as long as this choice is consistent between $\mathbfss{M}_\ell^{\alpha \beta}$ and $\widetilde{\mathbfit{a}}$.

\subsection{The joint distribution of quadratic forms}

Each cut-sky cross- or auto-spectrum estimator can be written in the form of Equation \eqref{Eqn:estimator_as_qf}, each with the same multivariate Gaussian vector $\widetilde{\mathbfit{a}}$ and different symmetric matrix $\mathbfss{M}_\ell^{\alpha \beta}$ depending on the field(s) and multipole in question. The joint characteristic function (CF) of an arbitrary number of such quadratic forms, $\varphi \left( \mathbfit{t} \right)$, has a known analytic form \citep{Good1963}:
\begin{equation}
    \varphi \left( \mathbfit{t} \right) =
    \Bigg\lvert \mathbfss{I} - 2i \sum_\ell \sum_{\alpha \beta}
    t_\ell^{\alpha \beta} \mathbfss{M}_\ell^{\alpha \beta} \bm{\Sigma}
    \Bigg\rvert^{-1/2},
    \label{Eqn:joint_cf}
\end{equation}
where $\lvert \cdot \rvert$ denotes the matrix determinant. The joint CF is defined as the expectation value $\langle e^{i \mathbfit{t} \cdot \widetilde{\mathbfit{C}}} \rangle$, which can also be written as a Fourier integral, i.e.
\begin{equation}
    \varphi \left( \mathbfit{t} \right) =
    \int d\widetilde{\mathbfit{C}}
    \exp \left( i \mathbfit{t} \cdot \widetilde{\mathbfit{C}} \right)
    f \left( \widetilde{\mathbfit{C}} \right),
    \label{Eqn:cf_fourier}
\end{equation}
where each $\left\{ \widetilde{C}_\ell^{\alpha \beta}, t_\ell^{\alpha \beta} \right\}$ are a Fourier pair\footnote{This should not be confused with the Fourier pair of $\left\{ \ell, \theta \right\}$ which relate the power spectrum $C_\ell$ to the correlation function $\xi \left( \theta \right)$.} and $\widetilde{\mathbfit{C}}$ and $\mathbfit{t}$ are the respective vectors of all $\widetilde{C}_\ell^{\alpha \beta}$ and $t_\ell^{\alpha \beta}$ for all fields and multipoles in question. $f \left( \widetilde{\mathbfit{C}} \right)$ is the joint probability density function of all observed pseudo-$C_\ell$s. Equation \eqref{Eqn:cf_fourier} may then be inverted to yield the exact joint distribution of the pseudo-$C_\ell$s:
\begin{equation}
    f \left( \widetilde{\mathbfit{C}} \right) =
    \frac{1}{\left( 2 \pi \right)^\nu}
    \int d\mathbfit{t}
    \exp \left( -i \widetilde{\mathbfit{C}} \cdot  \mathbfit{t} \right)
    \Bigg\lvert \mathbfss{I} - 2i \sum_\ell \sum_{\alpha \beta}
    t_{\ell}^{\alpha \beta} \mathbfss{M}_\ell^{\alpha \beta} \bm{\Sigma}
    \Bigg\rvert^{-1/2}.
    \label{Eqn:joint_likelihood}
\end{equation}
Here $\nu$ is the length of the data vector, which is equal to $N \times \left( \ell_\mathrm{max} + 1 \right)^2$ when considering $N$ fields and all multipoles from $0$ to $\ell_\mathrm{max}$. Equation \eqref{Eqn:joint_likelihood} is normalised by construction, such that it integrates to 1 over the range of possible $\widetilde{\mathbfit{C}}$.

Equation \eqref{Eqn:joint_likelihood} implicitly depends on the cosmological model via the covariance matrix of pseudo-$a_{\ell m}$s, $\bm{\Sigma}$, whose elements are given in Equations \eqref{Eqn:cov_re_re_general}--\eqref{Eqn:cov_re_im_general}. These elements depend on both the underlying power spectra and the mask for each field. When regarded as a function of the model parameters $\bm{\theta}$ for fixed observed pseudo-$C_\ell$ estimates, Equation \eqref{Eqn:joint_likelihood} is hence equivalent to the likelihood of the parameters, $\mathcal{L} \left( \bm{\theta} \right)$. Therefore, Equation \eqref{Eqn:joint_likelihood} is our second key result: the exact joint likelihood of an arbitrary number of auto- and cross-pseudo-$C_\ell$ estimates from an arbitrary number of correlated spin-0 and spin-2 fields. The posterior distribution of the parameters $p \left( \bm{\theta} \right)$ may then be obtained via Bayes' theorem, $p \left( \bm{\theta} \right) \propto \pi \left( \bm{\theta} \right) \mathcal{L} \left( \bm{\theta} \right)$, where $\pi \left( \bm{\theta} \right)$ is the prior distribution of the model parameters $\bm{\theta}$, and the proportionality constant is equal to the Bayesian evidence such that the posterior distribution is normalised.

We note the resemblance of Equation \eqref{Eqn:joint_likelihood} to Equation (16) of \cite{Wandelt2001} and to Equation (C6) of \cite{Hamimeche2008}, as all three are instances of the general distribution of quadratic forms. However, our Equation \eqref{Eqn:joint_likelihood} is a much more generally applicable result than Equation (16) of \cite{Wandelt2001}, which is restricted to a single multipole of a single spin-0 power spectrum observed with an azimuthally symmetric mask. Equation (C6) of \cite{Hamimeche2008}, meanwhile, describes the distribution of a single multipole of a full-sky weighted cross-spectrum estimator.

\subsubsection{The likelihood of a subset of pseudo-$C_\ell$ estimates}

In Equations \eqref{Eqn:a_hierarchy}--\eqref{Eqn:m_diag} we have presented a formalism for obtaining the full multivariate likelihood for all correlated fields and multipoles from $\ell_\mathrm{min}$ to $\ell_\mathrm{max}$. However, this likelihood may be adapted depending on exactly which pseudo-$C_\ell$s are required, and only the elements of $\bm{\Sigma}$ that contribute to the relevant estimators must be calculated. For example, if we wanted to calculate the joint likelihood of $\widetilde{C}_{\ell = 2}^{\alpha \alpha}$ and $\widetilde{C}_{\ell = 4}^{\alpha \alpha}$, we would have a pseudo-$a_{\ell m}$ vector of
\begin{equation}
    \widetilde{\mathbfit{a}} =
    \left(
    \Re \widetilde{a}_{2 0},\,
    \Re \widetilde{a}_{2 1},\,
    \Im \widetilde{a}_{2 1},\,
    \Re \widetilde{a}_{2 2},\,
    \Im \widetilde{a}_{2 2},\,
    \Re \widetilde{a}_{4 0},\,
    \Re \widetilde{a}_{4 1},\,
    \Im \widetilde{a}_{4 1},\,
    \Re \widetilde{a}_{4 2},\,
    \Im \widetilde{a}_{4 2},\,
    \Re \widetilde{a}_{4 3},\,
    \Im \widetilde{a}_{4 3},\,
    \Re \widetilde{a}_{4 4},\,
    \Im \widetilde{a}_{4 4}
    \right)^T,
\end{equation}
where we have omitted the $\left( \alpha \right)$ subscripts. We would then choose $\mathbfss{M}_2^{\alpha \alpha}$ and $\mathbfss{M}_4^{\alpha \alpha}$ to pick out the appropriate elements of $\widetilde{\mathbfit{a}}$.

It is mathematically straightforward to extend this formalism to obtain the exact distribution of deconvolved power estimates, as this is a linear operation in the pseudo-$C_\ell$s and hence a quadratic form in the pseudo-$a_{\ell m}$s. However, this approach offers no additional constraining power and introduces additional possible sources of error, so we choose not to discuss this further in this paper. Similarly, the extension to obtain the distribution of bandpowers through linear binning of multipoles---which is often necessary if one wishes to obtain deconvolved power estimates---is straightforward, requiring only appropriate changes to the selection matrices $\mathbfss{M}_\ell^{\alpha \beta}$.

\subsubsection{The likelihood of Quadratic Maximum Likelihood estimates}

At low $\ell$ one may prefer to use a Quadratic Maximum Likelihood (QML) power spectrum estimator \citep{Tegmark1997}, due to its optimality. The price is an increased computational cost, and diminishing returns at higher multipoles, compared to the pseudo-$C_\ell$ estimator \citep{Efstathiou2004}. By design, the QML estimator is also a quadratic form:
\begin{equation}
    y_\ell^{\alpha \beta} =
    \mathbfit{x}^T \mathbfss{E}_\ell^{\alpha \beta} \mathbfit{x}
    \quad \quad
    \text{with~}
    \mathbfss{E}_\ell^{\alpha \beta} = \frac{1}{2} \mathbfss{C}^{-1}
    \frac{\partial \mathbfss{C}}
    {\partial C_\ell^{\alpha \beta}} \mathbfss{C}^{-1},
    \label{Eqn:qml_estimator}
\end{equation}
where $\mathbfit{x}$ is the vector of all pixel values, which is multivariate Gaussian distributed with covariance \mathbfss{C}. The $y_\ell$ may be scaled and linearly transformed to provide unbiased estimates of the underlying power spectrum, but---as with pseudo-$C_\ell$ estimates---this step is unnecessary for cosmological inference and here we will only discuss the distribution of $y_\ell$ values.

From Equation \eqref{Eqn:qml_estimator} and the general distribution of quadratic forms, it is straightforward to write down the joint distribution of a set of QML estimates $\mathbfit{y}$,
\begin{equation}
    f \left( \mathbfit{y} \right) =
    \frac{1}{\left( 2 \pi \right)^\nu}
    \int d\mathbfit{t}
    \exp \left( -i \mathbfit{y} \cdot  \mathbfit{t} \right)
    \Bigg\lvert \mathbfss{I} - 2i \sum_\ell \sum_{\alpha \beta}
    t_{\ell}^{\alpha \beta} \mathbfss{E}_\ell^{\alpha \beta}
    \mathbfss{C}
    \Bigg\rvert^{-1/2},
    \label{Eqn:qml_distribution}
\end{equation}
where $\nu$ remains the length of the data vector, which is now equal to the total number of pixels across all maps. As with Equation \eqref{Eqn:joint_likelihood} for pseudo-$C_\ell$ estimates, Equation \eqref{Eqn:qml_distribution} gives the likelihood of model parameters $\mathcal{L} \left( \bm{ \theta } \right)$ when considered as a function of the model for fixed $\mathbfit{y}$. A potentially useful property of Equation \eqref{Eqn:qml_distribution} is that the matrices needed to evaluate it---$\mathbfss{C}$ and the set of $\mathbfss{E}_\ell^{\alpha \beta}$---are the same matrices needed to evaluate the estimator itself, reducing the amount of additional work needed to evaluate the likelihood.

Theoretically, one could also obtain the joint distribution of both pseudo-$C_\ell$ and QML estimates. This would require writing both as quadratic forms in the same underlying quantity, such as the full-sky $a_{\ell m}$s. This is mathematically possible because the transforms from full-sky $a_{\ell m}$s to both pixels and cut-sky pseudo-$a_{\ell m}$s are linear, meaning that a quadratic form in one can also be written as a quadratic form in the other. These transforms could then be encoded in the selection matrix in place of $\mathbfss{M}_\ell^{\alpha \beta}$ in Equation \eqref{Eqn:joint_likelihood} or $\mathbfss{E}_\ell^{\alpha \beta}$ in Equation \eqref{Eqn:qml_distribution}, with the comparatively simple full-sky $a_{\ell m}$ covariance matrix in place of $\bm{\Sigma}$ or $\mathbfss{C}$ respectively. However, in practice this is unlikely to be an attractive or computationally tractable option. An alternative would be to form the approximate joint distribution using a Gaussian copula, which would avoid the need to write both sets of estimates in terms of a common basis. Instead it would require the correlation coefficient or covariance between QML and pseudo-$C_\ell$ estimates, expressions for which are given in \cite{Efstathiou2004}.

\section{Application to Cosmic Microwave Background polarisation}
\label{Sec:examples}

In this section we will demonstrate that the pseudo-$C_\ell$ likelihood presented in Section \ref{Sec:pcl_dist} is correct by calculating the full joint distribution of three multipoles of the CMB polarisation power spectra, and comparing to the observed distribution obtained from simulations. We will show that our likelihood exactly reproduces correlations between spectra. It also naturally models correlations between any number of multipoles, but here we consider three-dimensional distributions only to limit the computational resources needed to calculate each full distribution.

\subsection[CMB pseudo-alm covariance]{CMB pseudo-$\widetilde{a}_{\ell m}$ covariance}
\label{Sec:CMB_alm_covariance}

In the CMB case, the relevant fields are the temperature field, the $E$-mode polarisation and the $B$-mode polarisation. We therefore identify
\begin{equation}
a_{\ell m}^{\left( \alpha \right)} \in
\left\{ T_{\ell m}, E_{\ell m}, B_{\ell m} \right\}; \quad \quad
\widetilde{a}_{\ell m}^{\left( \alpha \right)} \in
\left\{ \widetilde{T}_{\ell m},
\widetilde{E}_{\ell m},
\widetilde{B}_{\ell m} \right\}.
\end{equation}
The fields are mixed as \citep{Lewis2001, Brown2005}
\begin{equation}
\widetilde{T}_{\ell m} = \sum_{\ell' m'} {}_0W_{\ell \ell'}^{m m'} T_{\ell' m'};
\quad \quad
\widetilde{E}_{\ell m} = \sum_{\ell' m'} \left(
W_{\ell \ell' m m'}^+ E_{\ell' m'} + W_{\ell \ell' m m'}^- B_{\ell' m'}
\right);
\quad \quad
\widetilde{B}_{\ell m} = \sum_{\ell' m'} \left(
W_{\ell \ell' m m'}^+ B_{\ell' m'} - W_{\ell \ell' m m'}^- E_{\ell' m'}
\right),
\end{equation}
giving the relevant derivatives
\begin{equation}
\frac{ \partial \widetilde{T}_{\ell m} }{ \partial T_{\ell' m'} }
= {}_0W_{\ell \ell'}^{m m'}; \quad \quad
\frac{ \partial \widetilde{E}_{\ell m} }{ \partial E_{\ell' m'} }
= W_{\ell \ell' m m'}^+; \quad \quad
\frac{ \partial \widetilde{E}_{\ell m} }{ \partial B_{\ell' m'} }
= W_{\ell \ell' m m'}^-; \quad \quad
\frac{ \partial \widetilde{B}_{\ell m} }{ \partial E_{\ell' m'} }
= - W_{\ell \ell' m m'}^-; \quad \quad
\frac{ \partial \widetilde{B}_{\ell m} }{ \partial B_{\ell' m'} }
= W_{\ell \ell' m m'}^+.
\label{Eqn:cmb_derivatives}
\end{equation}
Here we consider only the polarisation fields, as CMB polarisation is the focus of many current and future experiments searching for evidence of inflation \citep[e.g.][]{hui2018, so2019, abazajian2016, hazumi2019}. However, our formalism naturally extends to include the temperature field, as well as including cross-correlation between fields observed by different detectors. Inserting the derivatives in Equation \eqref{Eqn:cmb_derivatives} into the general pseudo-$a_{\ell m}$ covariance matrix elements given in Equations \eqref{Eqn:cov_re_re_general}--\eqref{Eqn:cov_re_im_general} gives the elements of the CMB polarisation pseudo-$a_{\ell m}$ covariance matrix. Here we assume that all underlying $C_\ell^{EB}$ vanish, as is the case in any model in which parity is conserved, but Equations \eqref{Eqn:cov_re_re_general}--\eqref{Eqn:cov_re_im_general} naturally allow for non-zero underlying $C_\ell^{EB}$.

\subsection{Implementation}

The calculation of the pseudo-$a_{\ell m}$ covariance matrix requires theory power spectra and harmonic space window functions. We use \textsc{camb} \citep{Lewis2000, Howlett2012} through the \textsc{CosmoSIS}\footnote{https://bitbucket.org/joezuntz/cosmosis} interface \citep{Zuntz2015} to generate CMB power spectra using the $\Lambda$CDM$+r$ model with {\it Planck} 2018 best fit parameters from \citet{Planck2018VI} and tensor-to-scalar ratio $r = 0.01$. We manually implement the harmonic space window function transform in Equation \eqref{Eqn:harm_window} using spin-weighted spherical harmonics provided by the \texttt{spherical\_functions} Python package.\footnote{https://github.com/moble/spherical\_functions} However, we found that these only work reliably up to $\ell \approx 35$, so for our tests here we impose $\ell_\mathrm{max} = 30$. In future work, we intend to calculate the harmonic space window functions by expressing them in terms of Wigner $3j$ symbols \citep{Hivon2002} and making use of their recursion relations (\citealt{Lewis2001}; see also the Appendices of \citealt{Hamimeche2008, Hamimeche2009}).

When calculating the joint characteristic function in Equation \eqref{Eqn:joint_cf}, we find it is more numerically stable to use an alternative form that avoids the need to evaluate the determinant of a complex matrix:
\begin{equation}
    \varphi \left( \mathbfit{t} \right) =
    \prod_j \left(
    1 - 2i \lambda_j
    \right)^{-1/2},
    \quad \quad
    \lambda_j \in
    \lambda \left( \sum_\ell \sum_{\alpha \beta}
    t_\ell^{\alpha \beta} \mathbfss{M}_\ell^{\alpha \beta} \bm{\Sigma}
    \right);
    \label{Eqn:CF_alt}
\end{equation}
i.e., the product is over all eigenvalues of the real matrix
$\sum_\ell \sum_{\alpha \beta}
t_\ell^{\alpha \beta} \mathbfss{M}_\ell^{\alpha \beta} \bm{\Sigma}$. We show in Appendix \ref{App:cf_equivalance} that this form is mathematically equivalent to the form presented in Equation \eqref{Eqn:joint_cf}. To calculate the full joint likelihood distribution from the joint CF we implement Equation~\eqref{Eqn:joint_likelihood} by writing it in terms of a Fast Fourier Transform (FFT):
\begin{equation}
    f \left( \widetilde{\mathbfit{C}} \right) =
    \frac{1}{\left( 2 \pi \right)^n}
    \int_n \left( \prod_k dt_k \right)
    \exp \left( -i \sum_k
    \widetilde{C}_k t_k \right)
    \varphi \left( \mathbfit{t} \right)
    = \lim_{\Delta t_k \to 0}
    \frac{1}{\left( 2 \pi \right)^n}
    \left( \prod_k \Delta t_k \right)
    \exp \left( -i \sum_k t_{k0} \widetilde{C}_k \right)
    \mathrm{FFT}^n \big[ \varphi \left( \mathbfit{t} \right) \big],
\end{equation}
where we have used the index $k$ as a proxy to encapsulate all summation variables ($\ell$, $\alpha$, $\beta$) in Equation \eqref{Eqn:CF_alt}. We use the \texttt{numpy}\footnote{https://numpy.org} $n$-dimensional FFT \citep{Oliphant2007}:
\begin{equation}
\mathrm{FFT}^n \big[ f \left( \mathbfit{t} \right) \big] \left( \mathbfit{x} \right) =
\prod_k \sum_{m_k = 1}^{N_k}
f \left( \left\{ t_{k0} + m_k \Delta t_k \right\} \right)
\exp \left( -i \sum_{k'} \frac{ x_{k'} m_{k'} \Delta t_{k'} }{N_{k'}} \right),
\end{equation}
where each $t_k$ is discretised as $t_k = t_{k0} + m_k \Delta t_k$.

\subsection{Simulations}

We generated 276\,million realisations of the CMB polarisation field from the fixed theory power spectra using the \texttt{synfast} routine in {\tt HEALPix}\footnote{https://healpix.sourceforge.io} \citep{Gorski2005} with a resolution of \texttt{nside} $= 128$. To match the assumptions made in the theoretical distribution, we also impose $\ell_\text{max} = 30$ in the input power. We note that the accuracy of a finite $\ell_\text{max}$ requires either a band-limited signal (such as is the case for the CMB), a well-behaved mask or apodisation to strongly suppress long-range mode mixing. For each realisation, we applied the polarisation field mask used in the {\it WMAP} 9-year analysis, corresponding to a sky fraction of $73.2\,\%$. The mask is described further in \citet{Bennett2013}. We then measured the $EE$, $BB$ and $EB$ pseudo-$C_\ell$ power spectra from the masked maps using \texttt{anafast}. We formed marginal histograms of the power at $\ell = 2$, 5 and 10 from each spectrum and a three-dimensional histogram of the measured $\ell = 2$ power from all three spectra, which we selected to demonstrate the ability of our likelihood to exactly describe correlations between spectra. It also naturally (and exactly) describes correlations between multipoles, though we do not show this here to limit the computational complexity and number of required simulations. We have instead chosen to focus on the fact that our likelihood exactly models the mixing of power between $E$- and $B$-modes, as this is a particular difficulty for approximate methods due to the large discrepancy in the underlying $E$- and $B$-mode power. We have separately tested and confirmed the ability of our likelihood to exactly describe correlations between multipoles.

\subsection{Comparison to approximation}
\label{Eqn:sec_approx}

We also consider an approximation to our exact likelihood, to demonstrate the importance of accurately modelling the full multidimensional distribution. On the full sky, the joint distribution of $\widehat{C}_\ell^{EE}$, $\widehat{C}_\ell^{BB}$ and $\widehat{C}_\ell^{EB}$ for a single fixed multipole $\ell$ follows a Wishart distribution \citep[e.g.][]{Percival2006}:
\begin{equation}
\begin{pmatrix}
\widehat{C}_\ell^{EE} & \widehat{C}_\ell^{EB} \\
\widehat{C}_\ell^{EB} & \widehat{C}_\ell^{BB}
\end{pmatrix}
\sim
W_2 \left( \nu, \mathbfss{W}_\ell \right),
\quad \quad
\nu = 2 \ell + 1, \quad \quad
\mathbfss{W}_\ell =
\frac{1}{2 \ell + 1}
\begin{pmatrix}
C_\ell^{EE} & 0 \\
0 & C_\ell^{BB}
\end{pmatrix},
\end{equation}
where $W_p \left( d, \mathbfss{S} \right)$ denotes the rank-$p$ Wishart distribution with $d$ degrees of freedom and scale matrix $\mathbfss{S}$. We adapt this for the cut sky using ``effective" parameters of $\nu_\mathrm{eff}$ and $\mathbfss{W}_\ell^\mathrm{eff}$. We have chosen these effective parameters to be those that best fit the marginal distributions of $\widetilde{C}_\ell^{EE}$, $\widetilde{C}_\ell^{BB}$ and $\widetilde{C}_\ell^{EB}$, as calculated with our likelihood. We do not consider this to be a viable or attractive alternative to the exact likelihood; however, we include it to demonstrate that even an approximation that fits the marginals well fails to capture the full multidimensional likelihood. In addition to this, the Wishart approximation can only model correlations between different spectra for the same $\ell$, and not correlations between multipoles. Our likelihood, in contrast, exactly models correlations not only between spectra but also between multipoles, a behaviour which we have verified separately to the tests presented in this paper. However, as discussed above, we have chosen to focus on the former behaviour in this example.

A common alternative is to use a multivariate Gaussian approximation to the likelihood. With a suitable choice of mean, the problem then reduces to finding the most appropriate choice of covariance matrix, which can be calibrated with simulations. However, at low multipoles the true likelihood is so skewed that a Gaussian is an extremely poor approximation, which can lead to biased results and, if not modified, can assign non-zero probability to unphysical results such as negative auto-power spectra. For this reason, the 2018 {\it Planck} analysis uses a pixel-based likelihood below $\ell = 30$ as described in \citet{Planck2018V}.

\section{Results}
\label{Sec:results}

\subsection{Marginal distributions}

\begin{figure}
    \includegraphics[width=\columnwidth]{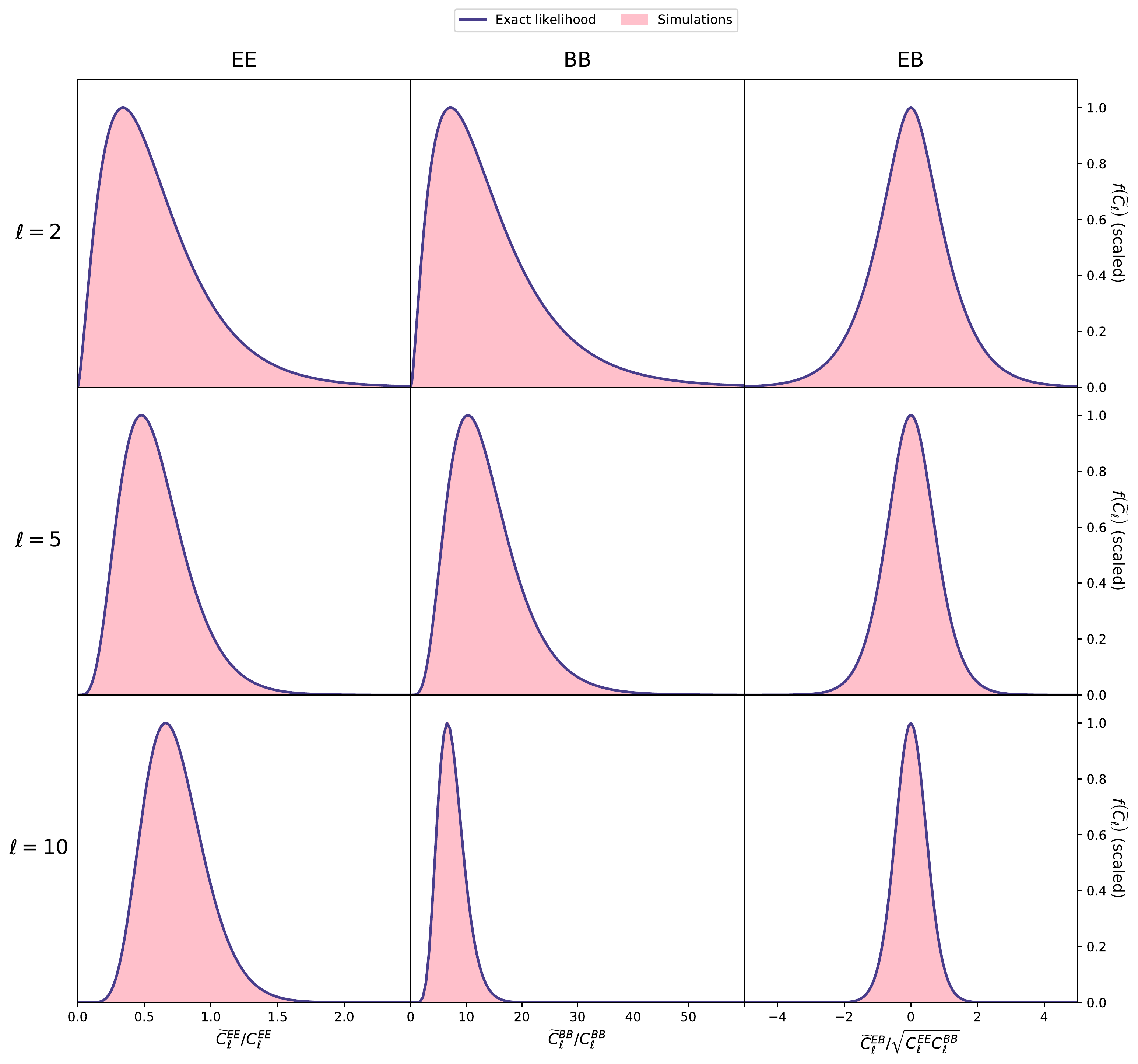}
    \caption{The marginal distributions of $\widetilde{C}_\ell^{EE}$, $\widetilde{C}_\ell^{BB}$ and $\widetilde{C}_\ell^{EB}$ for $\ell = 2$, 5 and 10, predicted by our exact likelihood (blue curves) compared to those observed in our simulations (pink histograms). The maximum value of each curve has been rescaled to 1 for ease of comparison.}
    \label{Fig:marginals}
\end{figure}

In Fig. \ref{Fig:marginals} we show the marginal distributions of $\widetilde{C}_\ell^{EE}$, $\widetilde{C}_\ell^{BB}$ and $\widetilde{C}_\ell^{EB}$ for each of $\ell = 2$, 5 and 10, to compare the prediction of our exact likelihood to the distributions observed in our simulations. Each histogram uses 300 bins. There is an excellent fit between the predicted and observed distribution, with no visible noise in the histograms due to the large number of events in each marginal distribution. The predicted likelihood exactly reproduces both the shape and amplitude of the observed distributions, including the considerable skewness in the auto-spectra. This skewness is reduced for higher multipoles, which is consistent with the full-sky behaviour of the likelihood. We have scaled each auto-$\widetilde{C}_\ell$ in Fig. \ref{Fig:marginals} by the relevant theory $C_\ell$ used to generate both the theoretical likelihood and the simulations. In the case of the cross-spectrum $\widetilde{C}_\ell^{EB}$, there is no input $C_\ell^{EB}$ so we instead normalise by $\sqrt{ C_\ell^{EE} C_\ell^{BB} }$. This scaling allows us to observe that the $E$-mode power is reduced by the sky cut while the $B$-mode power is increased. This is a result of $E$--$B$ mixing: the $EE$ power spectrum is much larger in magnitude than the $BB$ power spectrum (by a factor $\sim 200$ at $\ell = 2$), meaning that the $E$--$B$ mixing induced by the mask leads to a relative increase in $B$-mode power at the expense of $E$-mode power.

\subsection{Correlation between spectra}

\begin{figure}
    \includegraphics[width=\columnwidth]{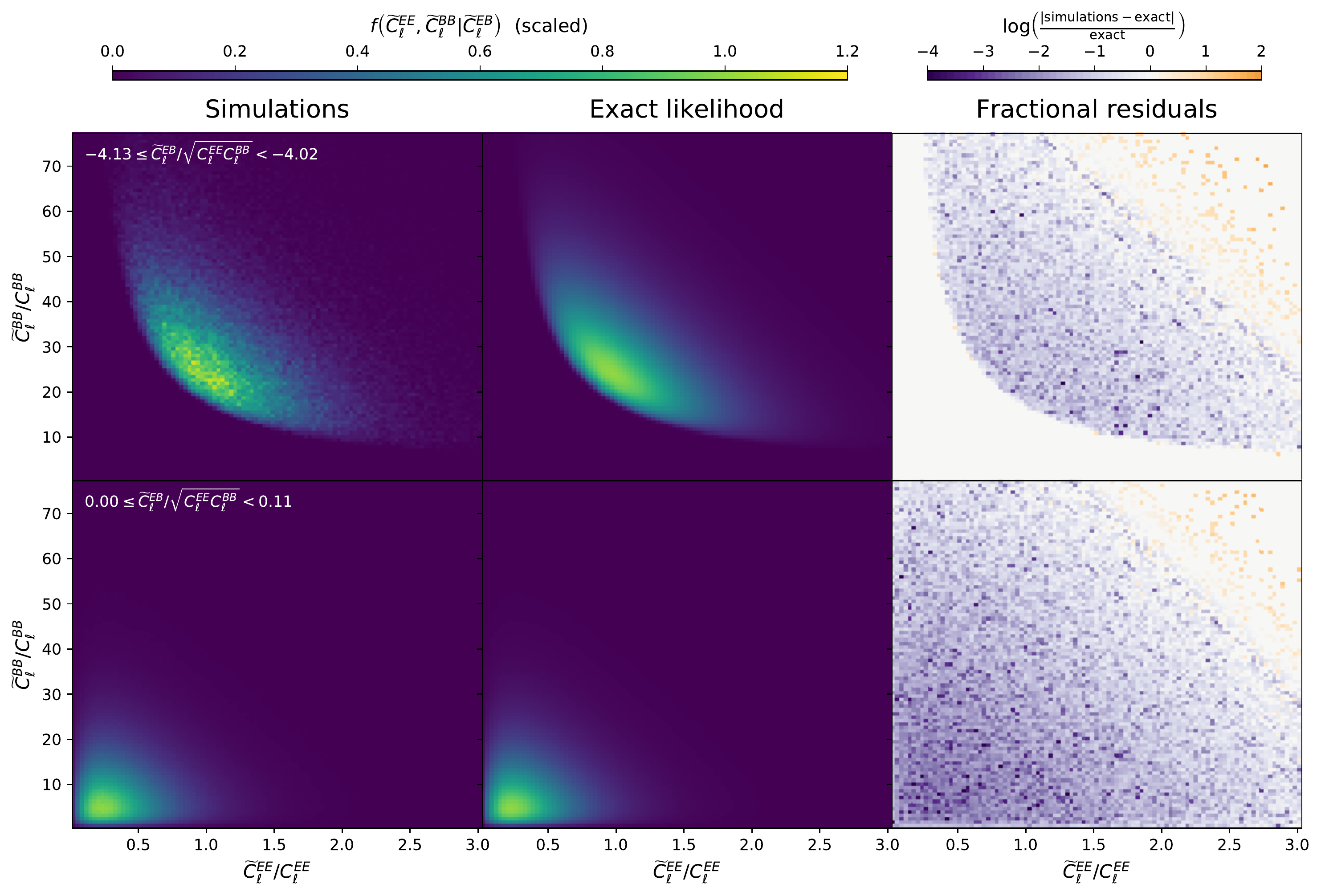}
    \caption{Two different slices of the joint $\widetilde{C}_\ell^{EE}$--$\widetilde{C}_\ell^{BB}$ distribution for $\ell = 2$, each for one fixed bin of $\widetilde{C}_\ell^{EB}$. The top row is the slice corresponding to $-4.13~\leq~\widetilde{C}_\ell^{EB} / \sqrt{ C_\ell^{EE} C_\ell^{BB} } < -4.02$ while the bottom row corresponds to $0.00~\leq~\widetilde{C}_\ell^{EB} / \sqrt{ C_\ell^{EE} C_\ell^{BB} } < 0.11$. The left panel in each row is the distribution observed from simulations, while the centre panel is the distribution predicted by our likelihood. The same colour scale is used for the left and centre panels within each row and has been chosen such that the exact likelihood in each slice runs between 0 and 1. The right panels show logarithmic fractional residuals, as defined in Equation \eqref{Eqn:resid_def}, and contain only noise due to the finite number of realisations.}
    \label{Fig:2D_EEBB}
\end{figure}
\begin{figure}
    \includegraphics[width=\columnwidth]{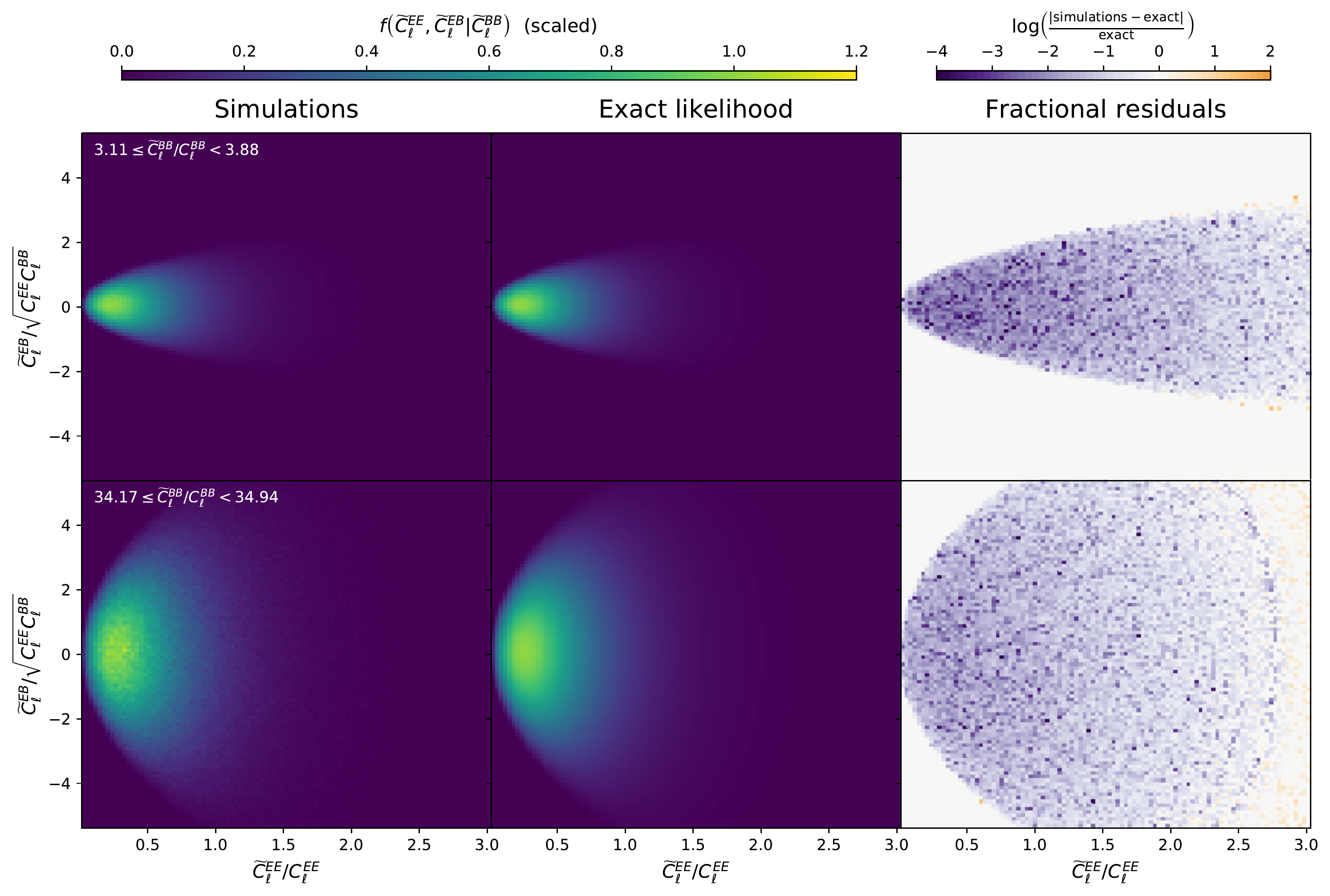}
    \caption{As Fig. \ref{Fig:2D_EEBB}, but for the $\widetilde{C}_\ell^{EE}$--$\widetilde{C}_\ell^{EB}$ distribution at fixed values of $\widetilde{C}_\ell^{BB}$. The top row corresponds to $3.11~\leq~\widetilde{C}_\ell^{BB} / C_\ell^{BB} < 3.88$ and the bottom row to $34.17~\leq~\widetilde{C}_\ell^{BB} / C_\ell^{BB} < 34.94$.}
\end{figure}
\begin{figure}
    \includegraphics[width=\columnwidth]{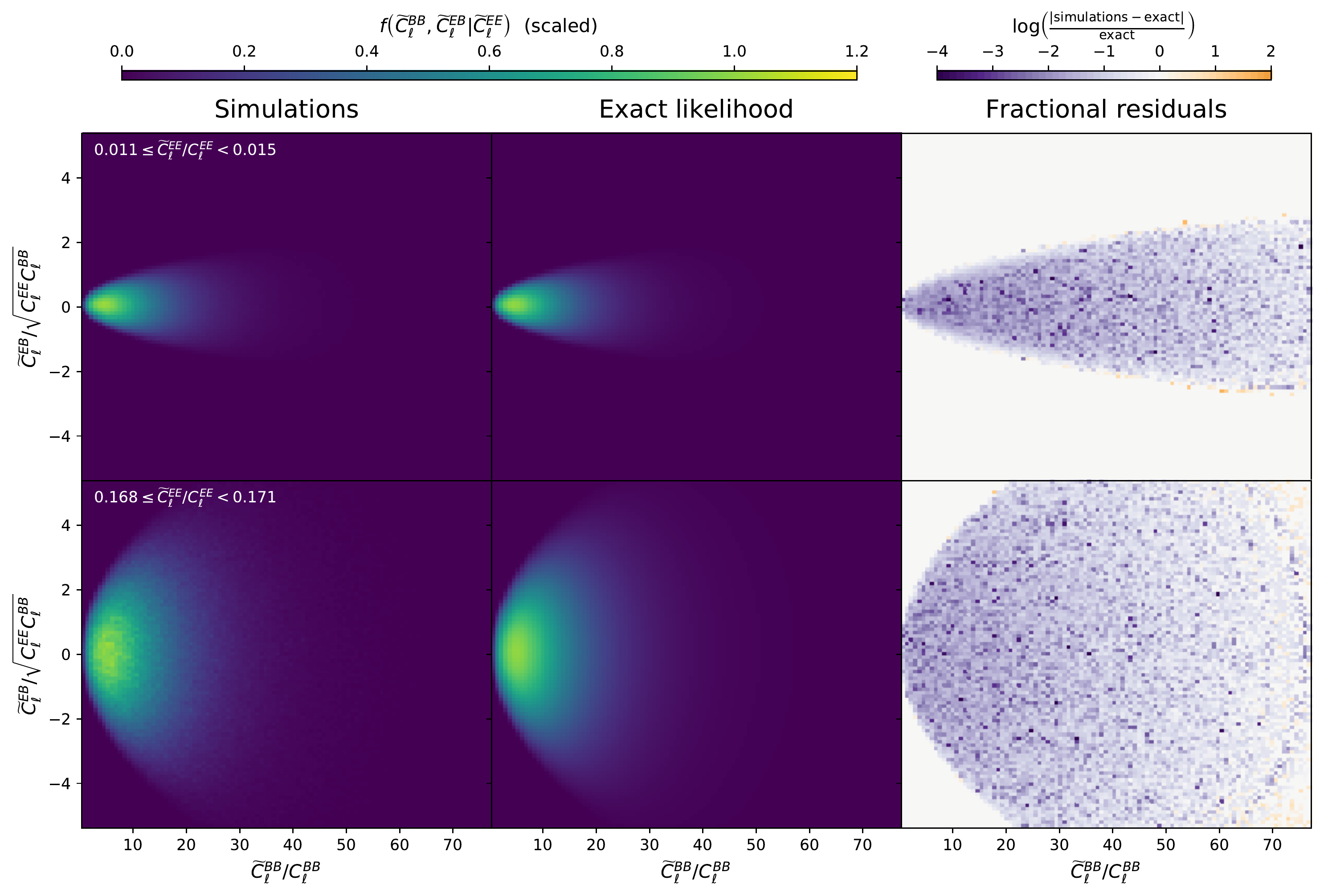}
    \caption{As Fig. \ref{Fig:2D_EEBB}, but for the $\widetilde{C}_\ell^{BB}$--$\widetilde{C}_\ell^{EB}$ distribution at fixed values of $\widetilde{C}_\ell^{EE}$. The top row corresponds to $0.011~\leq~\widetilde{C}_\ell^{EE} / C_\ell^{EE} < 0.015$ and the bottom row to $0.168~\leq~\widetilde{C}_\ell^{EE} / C_\ell^{EE} < 0.171$.}
    \label{Fig:2D_BBEB}
\end{figure}

As well as exactly reproducing marginal distributions, our likelihood naturally describes correlations both between multipoles of the same spectrum and between spectra. As described in Section \ref{Sec:examples}, we have formed the three-dimensional joint likelihood of $\widetilde{C}_\ell^{EE}$, $\widetilde{C}_\ell^{BB}$ and $\widetilde{C}_\ell^{EB}$ for $\ell = 2$. We formed the corresponding simulated distribution by binning events in three dimensions, using 100 bins in each dimension. We then integrated our exact likelihood over the volume of each histogram bin to allow for comparison between theory and simulations. In Figs. \ref{Fig:2D_EEBB}--\ref{Fig:2D_BBEB} we show two-dimensional slices of this three-dimensional likelihood. Each slice corresponds to fixing a single histogram bin in one dimension and shows all bins in the other two dimensions. Our likelihood appears to accurately match the observed distributions in all six slices to within pixel noise that arises from the finite number of realisations in the simulations. The right-hand panel for each slice shows the logarithmic fractional residual, defined as
\begin{equation}
    r = \log_{10} \left(
    \frac{\big\lvert
    \text{sampled density from simulations}
    - \text{density from exact likelihood}
    \big\rvert}{\text{density from exact likelihood}}
    \right).
    \label{Eqn:resid_def}
\end{equation}
Bins with no sampled events have $r = 0$ and appear as white in Figs. \ref{Fig:2D_EEBB}--\ref{Fig:2D_BBEB}. These areas were not explicitly excluded, but their probability is very low (albeit non-zero). We do not otherwise see any clear evidence of structure in these residuals, indicating that these bins contain only noise. This is mostly at the level of $r \approx -4$ to $-2$, except for a small number of outlying bins whose probability density is so low that the expected number of events in each bin from our 276\,million simulations is significantly less than 1, leading to fractional residual values up to $r \approx 2$ in those bins in which an event was observed.

\subsubsection{Comparison to approximation}

\begin{figure}
    \includegraphics[width=\columnwidth]{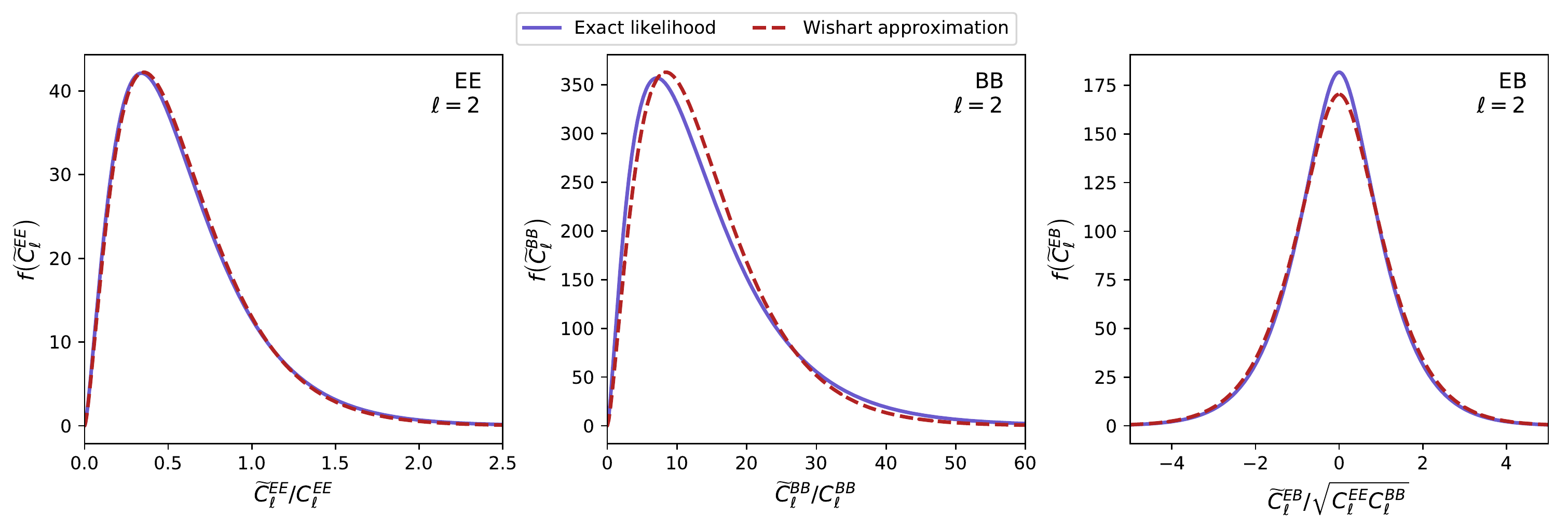}
    \caption{Marginal distributions based on the Wishart approximation with best-fitting parameters given in Equation \eqref{Eqn:wish_params} (dashed red curves) compared to the exact likelihood (solid blue curves).}
    \label{Fig:marg_vs_approx}
\end{figure}

In this section we compare to a Wishart distribution with fitted parameters $\nu_\mathrm{eff}$ and $\mathbfss{W}_\ell^\mathrm{eff}$, as described in Section \ref{Eqn:sec_approx}. We are not advocating the use of this approximation; on the contrary, we aim to demonstrate the merits of using our exact likelihood. For this reason, we are not concerned with whether we can obtain appropriate values of $\nu_\mathrm{eff}$ and $\mathbfss{W}_\ell^\mathrm{eff}$ in practice. We simultaneously fitted the three marginals of a $p=2$ Wishart distribution to the exact marginals, and obtained the following best-fitting values:
\begin{equation}
    \frac{\nu_\mathrm{eff}}{2 \ell + 1} = 1.0; \quad \quad
    \mathbfss{W}_\ell^\mathrm{eff} = \frac{1}{2 \ell + 1}
    \begin{pmatrix}
    0.59 C_\ell^{EE} & 0 \\
    0 & 14 C_\ell^{BB}
    \end{pmatrix}.
    \label{Eqn:wish_params}
\end{equation}
We have no reason to expect that these values would also be the best-fitting values for a different $\ell$ or for a different input cosmology, and they would certainly be different for another mask. We show the resulting marginal distributions in Fig. \ref{Fig:marg_vs_approx}, where we have omitted the simulated histogram for clarity. The fit is almost perfect for $\widetilde{C}_\ell^{EE}$, indicating that this marginal distribution closely follows a gamma distribution as in the full-sky case. There is slightly more deviation for $\widetilde{C}_\ell^{BB}$ and $\widetilde{C}_\ell^{EB}$.

\begin{figure}
    \includegraphics[width=\columnwidth]{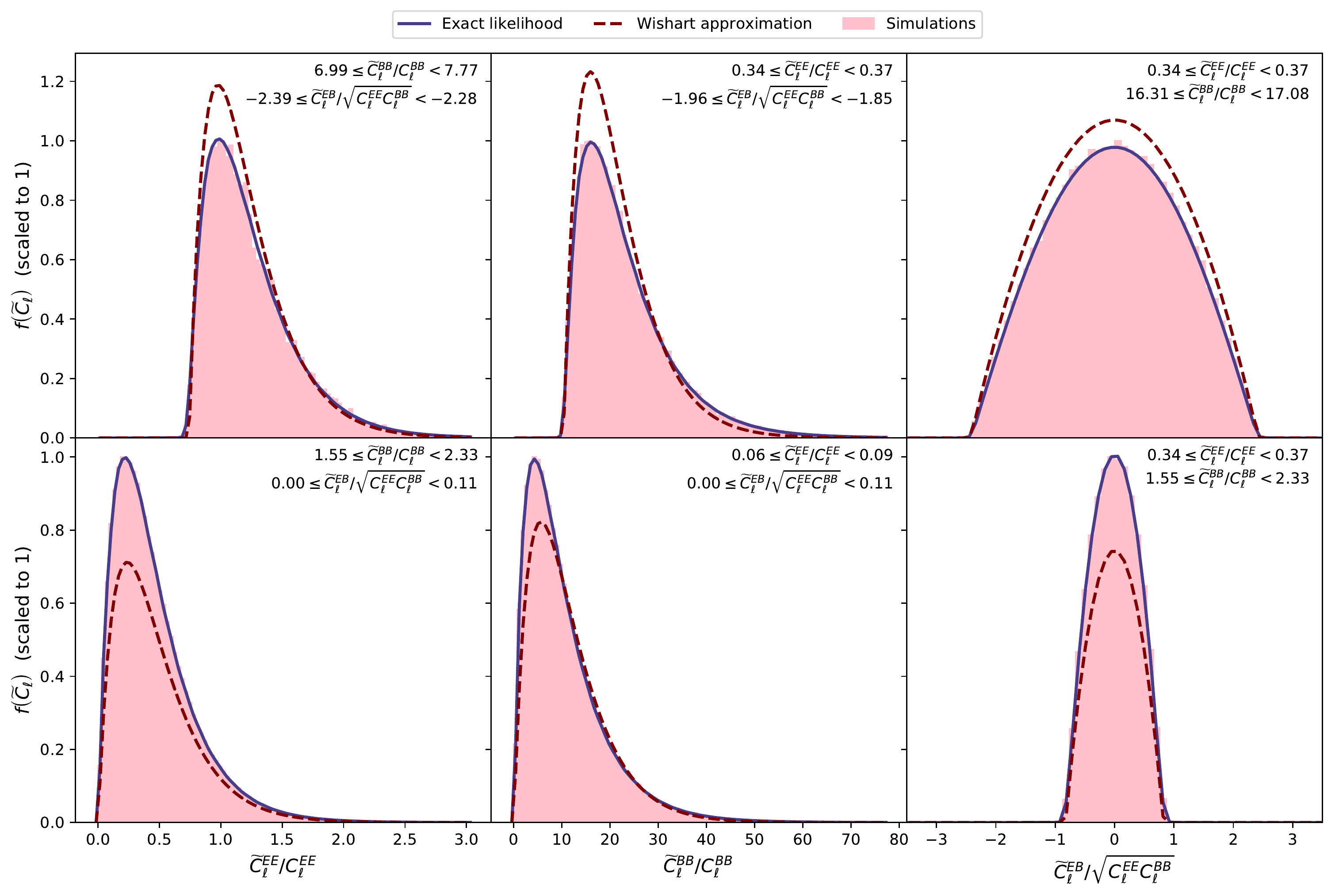}
    \caption{One-dimensional slices of the joint distribution of $\widetilde{C}_\ell^{EE}$, $\widetilde{C}_\ell^{BB}$ and $\widetilde{C}_\ell^{EB}$ for $\ell = 2$. In each slice, two of the values are fixed while the third is allowed to vary.}
    \label{Fig:1dslices_vs_approx}
\end{figure}

\begin{figure}
    \includegraphics[width=\columnwidth]{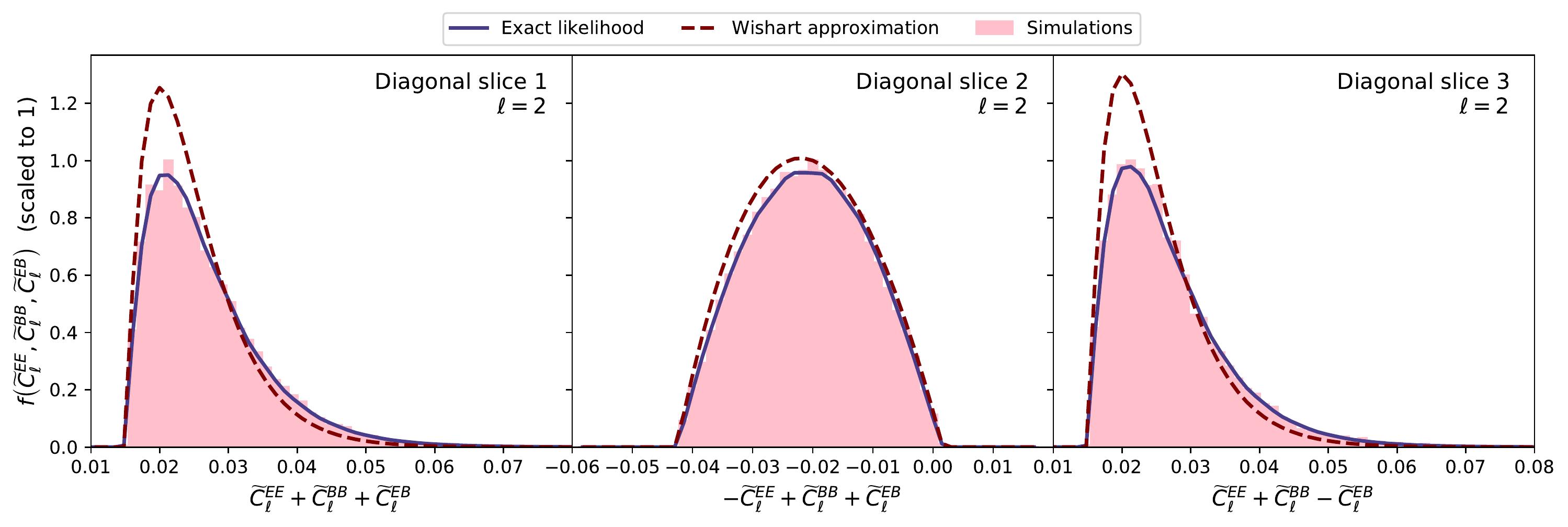}
    \caption{As Fig. \ref{Fig:1dslices_vs_approx} but for slices taken in three diagonal directions, each corresponding to a linear combination of the three pseudo-$C_\ell$s.}
    \label{Fig:diag_vs_approx}
\end{figure}

We integrated the Wishart probability density over each histogram bin in three-dimensional space, as we did with our exact likelihood. In Fig. \ref{Fig:1dslices_vs_approx} we show one-dimensional slices through this three-dimensional likelihood. For each slice, we have fixed a single bin for two dimensions and show the distribution across the third dimension. We see that while the marginal distributions suggest a near-exact fit for the Wishart approximation, the one-dimensional slices reveal that the approximation incorrectly distributes probability in some parts of the three-dimensional space relative to other parts. Our exact likelihood, on the other hand, faithfully reproduces the observed distribution throughout. This is also seen in Fig. \ref{Fig:diag_vs_approx}, which shows three one-dimensional slices in different diagonal directions across the three-dimensional space. In each of these slices, the Wishart approximation overestimates the probability density relative to the true distribution, implying that it must underestimate it in other parts of the distribution, given that the whole distribution is normalised to integrate to 1.

As well as failing to accurately reproduce the full distribution for a fixed $\ell$, the Wishart distribution cannot naturally be extended to include correlations between multipoles, whereas our exact likelihood automatically produces the full joint distribution between all multipoles of all power spectra. In some cases an approximation may work better over the joint distribution of many multipoles than for a single $\ell$ \citep{Hamimeche2008}, but the inverse can also be true \citep{Elsner2012}. In any case, the choice of approximation for the purposes of this comparison is unimportant: no approximation can completely match the exact distribution.

\section{Conclusions}
\label{Sec:conclusions}

We have presented the exact joint likelihood of an arbitrary number of pseudo-$C_\ell$ estimates from correlated Gaussian fields, valid for both auto- and cross-power spectra and for any mask geometry. The likelihood---given in Equation \eqref{Eqn:joint_likelihood}---naturally models both intrinsic correlations between spin-0 and spin-2 fields, and correlations induced by a cut sky which result in the mixing between spherical harmonic coefficients. The pseudo-$a_{\ell m}$s follow a multivariate Gaussian distribution with covariance matrix elements given by Equations \eqref{Eqn:cov_re_re_general}--\eqref{Eqn:cov_re_im_general}. In Equation \eqref{Eqn:qml_distribution} we have also presented the exact joint likelihood for QML power spectrum estimates. An accurate likelihood function is an essential companion to any estimator for unbiased cosmological inference, but until now a complete likelihood for either the pseudo-$C_\ell$ or QML estimator on an arbitrary sky has not been known.

We have shown how our likelihood can be applied to observations of the polarisation of the Cosmic Microwave Background. This is especially relevant given current and future experiments aimed at detecting primordial $B$-modes, which require exquisite control of all possible sources of systematic bias. One such source of bias is an inexact likelihood function, so knowledge of the exact likelihood could play an important role in extracting cosmological information from polarisation measurements in an unbiased manner. In particular, it exactly models the leakage of $E$-mode power into the much smaller $B$-mode signal. Our likelihood also extends naturally to include correlations between temperature anisotropy and polarisation, including cross-correlations between any number of detectors, where each observed temperature or polarisation field may have its own mask. It does not account for weak gravitational lensing of the CMB, which breaks the assumption of Gaussianity at higher multipoles.

The exact likelihood will also be extremely useful for weak lensing observations. It will perhaps be most valuable at relatively low multipoles, as this is the regime where the common assumption of a Gaussian likelihood for power spectrum estimates is least applicable due to the considerable skewness in the true likelihood \citep[e.g.][]{Sellentin2018a}. These low multipoles correspond to large physical scales, for which it is an excellent approximation to describe the spin-2 cosmic shear field as a Gaussian field. At higher multipoles there will be significant deviations from Gaussianity, so our likelihood cannot be considered exact on small scales. The likelihood naturally extends to describe the full distribution of auto- and cross-power spectra between an arbitrary number of redshift bins, each with its own mask. It is thus well suited for extracting robust cosmological constraints from tomographic galaxy clustering and weak lensing shear power spectrum measurements in multiple redshift bins, such as those that will result from future analyses of the {\it Euclid} \citep{Laureijs2011}, LSST \citep{LSST2009} and SKA \citep{SKA2018} surveys.

Use of the exact pseudo-$C_\ell$ likelihood is likely to be competitive in terms of speed when considering a small number of bandpower estimates. Its main strength compared to an exact pixel-based likelihood is that for a single power estimate the pseudo-$C_\ell$ likelihood requires the determinant evaluation of a matrix of size $\sim \ell$ compared to $\sim \ell^2$ for a pixel-based method. This means that it may be evaluated at much higher $\ell$ than is possible for a pixel-based method. However, once many power estimates are considered the scaling is less competitive. The pixel-based method would offer all additional multipoles $\ell' < \ell$ without significant additional computational cost, while the exact pseudo-$C_\ell$ likelihood scales as $\sim k^N$, where $N$ is the number of (band)power estimates and $k \sim 200$. The cost is driven by the need to evaluate the characteristic function in Equation \eqref{Eqn:joint_cf} for every value of the vector $\mathbfit{t}$. The range of $\mathbfit{t}$ must cover a wide enough space for the integral in the likelihood expression to converge, while at a sufficiently high resolution so that its curvature is accurately represented. Each point in $\mathbfit{t}$-space carries its own determinant or eigenvalue calculation (which is not the case for a single power estimate, due to the simple scaling of the determinant of a single matrix). We therefore recommend use of the exact likelihood only in the case of a small number of bandpower estimates, and are exploring alternative (necessarily approximate) approaches for the joint distribution of many power estimates, including the use of a copula with the exact marginal distributions. Copula methods have previously been described in a cosmological context, but only with approximate marginal distributions \citep{Benabed2009, Sato2010, Sato2011}. Alternative approaches previously explored in the literature include approximate extensions to the Wishart distribution to model correlations between multipoles \citep{Hamimeche2008, Mangilli2015}. Computational limitations in the likelihood calculation may also be mitigated to some extent by potential speed increases at other levels in the inference process such as neural net-assisted sampling \citep{Manrique-Yus2019}.

Despite the limitations of its direct use, knowledge of the exact pseudo-$C_\ell$ and QML likelihood is extremely useful as a starting point and testing benchmark for developing fast, accurate approximations. The current best practice for cosmological inference \citep[e.g.][]{Planck2018V} is to use an exact pixel-based likelihood at low multipoles and to switch to an approximate Gaussian power spectrum likelihood for higher multipoles, at the point at which a pixel-based likelihood becomes computationally unfeasible (at $\ell = 29$ in the case of {\it Planck}, while for weak lensing analyses with many redshift bins an exact pixel-based method may not be feasible at all). Methods derived from this exact likelihood may fill an important niche between these two regimes, allowing the use of an exact or near-exact likelihood up to higher multipoles than is currently possible. This may be a powerful tool for interpreting future observations, given the increased statistical precision that they will offer. Our likelihood also has the advantage that it can naturally describe the cross-correlation power spectrum measured between two different maps, in contrast to exact pixel-based methods which are not readily adapted to extracting just the cross-correlation information. Considering only cross-spectra in this way makes the cosmological analysis insensitive to the details of the noise bias, which will be especially relevant for cosmic shear observations, for which shape noise is an important and uncertain factor.

More work is required to adapt our techniques so that they are useful in practice. In future work, we plan to investigate the feasibility of implementing approximations to the exact likelihood as part of a realistic cosmological inference pipeline, and to benchmark their performance against alternatives such as an exact pixel-based likelihood. It also remains to be seen up to what maximum multipole these methods can be applied---both computationally, and also regarding the effects of non-Gaussianity at higher multipoles. A cut sky mixes power between multipoles, so even relatively low $\ell$ modes may be contaminated by non-Gaussian power in the presence of a severe mask. We plan to assess the impact of this, and possibly even incorporate it into a more advanced weak lensing likelihood. We have also not considered noise or systematic errors at this stage.

\section*{Acknowledgements}


We thank Andrew Jaffe, Benjamin Joachimi and Andy Taylor for useful discussions, and the referee for helpful comments that improved the manuscript. REU is supported by a studentship from the UK Science and Technology Facilities Council. LW is supported by a UK Space Agency grant. Some of the results in this paper have been derived using the {\tt HEALPix} \citep{Gorski2005} package.




\bibliographystyle{mnras}
\bibliography{Mendeley}





\appendix

\section[General pseudo-alm covariance derivation]{General pseudo-$\lowercase{a}_{\ell \lowercase{m}}$ covariance derivation}
\label{App:cov_derivation}

In this appendix we derive the general covariance matrix elements of the pseudo-$a_{\ell m}$s. First we present the derivation of the real and imaginary parts of the general pseudo-$a_{\ell m}$s given in Equations \eqref{Eqn:Re_alm_general} and \eqref{Eqn:Im_alm_general}. We begin with the general form of a pseudo-$a_{\ell m}$ as a weighted sum of full-sky $a_{\ell m}s$ as given in Equation \eqref{Eqn:pseudo_alm_as_sum}. We expand the sum over $m'$ into separate sums for $m' < 0$ and $m' > 0$, and a term for $m' = 0$:
\begin{equation}
\widetilde{a}_{\ell m}^{ \left( \alpha \right) } =
\sum_{\beta, \ell'} \left[
\frac{\partial \widetilde{a}_{\ell m}^{ \left( \alpha \right) }}
{\partial a_{\ell' 0}^{ \left( \beta \right) }}
a_{\ell' 0}^{ \left( \beta \right) }
+ \sum_{m' < 0}
\frac{\partial \widetilde{a}_{\ell m}^{ \left( \alpha \right) }}
{\partial a_{\ell' m'}^{ \left( \beta \right) }}
a_{\ell' m'}^{ \left( \beta \right) }
+ \sum_{m' > 0}
\frac{\partial \widetilde{a}_{\ell m}^{ \left( \alpha \right) }}
{\partial a_{\ell' m'}^{ \left( \beta \right) }}
a_{\ell' m'}^{ \left( \beta \right) }
\right].
\end{equation}
Writing the $m' < 0$ contributions in terms of their $m' > 0$ counterparts using Equation \eqref{Eqn:alm_symmetry} gives
\begin{equation}
\begin{split}
\widetilde{a}_{\ell m}^{ \left( \alpha \right) } &=
\sum_{\beta, \ell'} \left[
\frac{\partial \widetilde{a}_{\ell m}^{ \left( \alpha \right) }}
{\partial a_{\ell' 0}^{ \left( \beta \right) }}
a_{\ell' 0}^{ \left( \beta \right) }
+ \sum_{m' < 0}
\frac{\partial \widetilde{a}_{\ell m}^{ \left( \alpha \right) }}
{\partial a_{\ell' m'}^{ \left( \beta \right) }}
\left( -1 \right)^{m'}
\left( a_{\ell' \lvert m' \rvert}^{\left( \beta \right)} \right)^*
+ \sum_{m' > 0}
\frac{\partial \widetilde{a}_{\ell m}^{ \left( \alpha \right) }}
{\partial a_{\ell' m'}^{ \left( \beta \right) }}
a_{\ell' m'}^{ \left( \beta \right) }
\right] \\
&= \sum_{\beta, \ell'} \left[
\frac{\partial \widetilde{a}_{\ell m}^{ \left( \alpha \right) }}
{\partial a_{\ell' 0}^{ \left( \beta \right) }}
a_{\ell' 0}^{ \left( \beta \right) }
+ \sum_{m' > 0} \left(
\frac{\partial \widetilde{a}_{\ell m}^{ \left( \alpha \right) }}
{\partial a_{\ell' m'}^{ \left( \beta \right) }}
a_{\ell' m'}^{ \left( \beta \right) }
+ \left( -1 \right)^{m'}
\frac{\partial \widetilde{a}_{\ell m}^{ \left( \alpha \right) }}
{\partial a_{\ell' -m'}^{ \left( \beta \right) }}
\left( a_{\ell' m'}^{ \left( \beta \right) } \right)^*
\right)
\right].
\end{split}
\end{equation}
We now expand this into real and imaginary parts as
\begin{equation}
\begin{split}
\widetilde{a}_{\ell m}^{ \left( \alpha \right) } =
\sum_{\beta, \ell'} \Bigg[ &
\left[ \Re \left(
\frac{\partial \widetilde{a}_{\ell m}^{ \left( \alpha \right) }}
{\partial a_{\ell' 0}^{ \left( \beta \right) }}
\right) + i \Im \left(
\frac{\partial \widetilde{a}_{\ell m}^{ \left( \alpha \right) }}
{\partial a_{\ell' 0}^{ \left( \beta \right) }}
\right) \right]
\Re \left( a_{\ell' 0}^{ \left( \beta \right) } \right) \\
& +
\sum_{m' > 0} \Bigg(
\left[ \Re \left(
\frac{\partial \widetilde{a}_{\ell m}^{ \left( \alpha \right) }}
{\partial a_{\ell' m'}^{ \left( \beta \right) }}
\right) + i \Im \left(
\frac{\partial \widetilde{a}_{\ell m}^{ \left( \alpha \right) }}
{\partial a_{\ell' m'}^{ \left( \beta \right) }}
\right) \right]
\left[ \Re \left(
a_{\ell' m'}^{ \left( \beta \right) }
\right) + i \Im \left(
a_{\ell' m'}^{ \left( \beta \right) }
\right) \right] \\
&~~~~~~~~~~~~~~ +
\left( -1 \right)^{m'}
\left[ \Re \left(
\frac{\partial \widetilde{a}_{\ell m}^{ \left( \alpha \right) }}
{\partial a_{\ell' -m'}^{ \left( \beta \right) }}
\right) + i \Im \left(
\frac{\partial \widetilde{a}_{\ell m}^{ \left( \alpha \right) }}
{\partial a_{\ell' -m'}^{ \left( \beta \right) }}
\right) \right]
\left[ \Re \left(
a_{\ell' m'}^{ \left( \beta \right) }
\right) - i \Im \left(
a_{\ell' m'}^{ \left( \beta \right) }
\right) \right]
\Bigg) \Bigg],
\end{split}
\end{equation}
which, after some algebra, gives the real and imaginary parts in Equations \eqref{Eqn:Re_alm_general} and \eqref{Eqn:Im_alm_general}.

We now present the derivation of the covariance between two of the real parts, as given in Equation \eqref{Eqn:cov_re_re_general}. The derivations of the imaginary--imaginary and real--imaginary covariance given in Equations \eqref{Eqn:cov_im_im_general} and \eqref{Eqn:cov_re_im_general} follow analogously. We begin by inserting the expression for the real part of a pseudo-$a_{\ell m}$ from Equation \eqref{Eqn:Re_alm_general}:
\begin{equation}
\begin{split}
& \mathrm{Cov} \left(
\Re \left( \widetilde{a}_{\ell m}^{ \left( \alpha \right) } \right),
\Re \left( \widetilde{a}_{\ell' m'}^{ \left( \beta \right) } \right)
\right) \\
&=
\mathrm{Cov} \Bigg(
\sum_{\gamma, \ell''} \Bigg[
\Re \left(
\frac{\partial \widetilde{a}_{\ell m}^{ \left( \alpha \right) }}
{\partial a_{\ell'' 0}^{ \left( \gamma \right) }}
\right)
\Re \left( a_{\ell'' 0}^{ \left( \gamma \right) } \right) \\
&~~~~~~~~~~~~~ +
\sum_{m'' > 0} \Bigg( \left[
\Re \left(
\frac{\partial \widetilde{a}_{\ell m}^{ \left( \alpha \right) }}
{\partial a_{\ell'' m''}^{ \left( \gamma \right) }}
\right)
+ \left( -1 \right)^{m''}
\Re \left(
\frac{\partial \widetilde{a}_{\ell m}^{ \left( \alpha \right) }}
{\partial a_{\ell'' -m''}^{ \left( \gamma \right) }}
\right) \right]
\Re \left(
a_{\ell'' m''}^{ \left( \gamma \right) }
\right) -
\left[ \Im \left(
\frac{\partial \widetilde{a}_{\ell m}^{ \left( \alpha \right) }}
{\partial a_{\ell'' m''}^{ \left( \gamma \right) }}
\right)
- \left( -1 \right)^{m''}
\Im \left(
\frac{\partial \widetilde{a}_{\ell m}^{ \left( \alpha \right) }}
{\partial a_{\ell'' -m''}^{ \left( \gamma \right) }}
\right) \right]
\Im \left(
a_{\ell'' m''}^{ \left( \gamma \right) }
\right) \Bigg) \Bigg], \\
&~~~~~~~~~~
\sum_{\varepsilon, \ell'''} \Bigg[
\Re \left(
\frac{\partial \widetilde{a}_{\ell' m'}^{ \left( \beta \right) }}
{\partial a_{\ell''' 0}^{ \left( \varepsilon \right) }}
\right)
\Re \left( a_{\ell''' 0}^{ \left( \varepsilon \right) } \right) \\
&~~~~~~~~~~~~~ +
\sum_{m''' > 0} \Bigg( \left[
\Re \left(
\frac{\partial \widetilde{a}_{\ell' m'}^{ \left( \beta \right) }}
{\partial a_{\ell''' m'''}^{ \left( \varepsilon \right) }}
\right)
+ \left( -1 \right)^{m'''}
\Re \left(
\frac{\partial \widetilde{a}_{\ell' m'}^{ \left( \beta \right) }}
{\partial a_{\ell''' -m'''}^{ \left( \varepsilon \right) }}
\right) \right]
\Re \left(
a_{\ell''' m'''}^{ \left( \varepsilon \right) }
\right) -
\left[ \Im \left(
\frac{\partial \widetilde{a}_{\ell' m'}^{ \left( \beta \right) }}
{\partial a_{\ell''' m'''}^{ \left( \varepsilon \right) }}
\right)
- \left( -1 \right)^{m'''}
\Im \left(
\frac{\partial \widetilde{a}_{\ell' m'}^{ \left( \beta \right) }}
{\partial a_{\ell''' -m'''}^{ \left( \varepsilon \right) }}
\right) \right]
\Im \left(
a_{\ell''' m'''}^{ \left( \varepsilon \right) }
\right) \Bigg) \Bigg] \Bigg).
\end{split}
\end{equation}
We expand this into a linear sum of full-sky $a_{\ell m}$ covariances, noting that the terms involving covariance between real and imaginary full-sky $a_{\ell m}$s immediately vanish, as do terms involving covariance between $m = 0$ and $m > 0$ $a_{\ell m}$s:
\begin{equation}
\begin{split}
&
\mathrm{Cov} \left(
\Re \left( \widetilde{a}_{\ell m}^{ \left( \alpha \right) } \right),
\Re \left( \widetilde{a}_{\ell' m'}^{ \left( \beta \right) } \right)
\right) \\
&= \sum_{\gamma, \varepsilon}
\sum_{\ell'' \ell'''}
\Bigg[
\Re \left(
\frac{\partial \widetilde{a}_{\ell m}^{ \left( \alpha \right) }}
{\partial a_{\ell'' 0}^{ \left( \gamma \right) }}
\right)
\Re \left(
\frac{\partial \widetilde{a}_{\ell' m'}^{ \left( \beta \right) }}
{\partial a_{\ell''' 0}^{ \left( \varepsilon \right) }}
\right)
\mathrm{Cov} \left(
\Re \left( a_{\ell'' 0}^{ \left( \gamma \right) } \right),
\Re \left( a_{\ell''' 0}^{ \left( \varepsilon \right) } \right)
\right) \\
&~~~~~~~~~~~~~~~~~~~~ +
\sum_{\substack{m'' > 0 \\ m''' > 0}} \Bigg(
\left[
\Re \left(
\frac{\partial \widetilde{a}_{\ell m}^{ \left( \alpha \right) }}
{\partial a_{\ell'' m''}^{ \left( \gamma \right) }}
\right)
+ \left( -1 \right)^{m''}
\Re \left(
\frac{\partial \widetilde{a}_{\ell m}^{ \left( \alpha \right) }}
{\partial a_{\ell'' -m''}^{ \left( \gamma \right) }}
\right) \right]
\left[
\Re \left(
\frac{\partial \widetilde{a}_{\ell' m'}^{ \left( \beta \right) }}
{\partial a_{\ell''' m'''}^{ \left( \varepsilon \right) }}
\right)
+ \left( -1 \right)^{m'''}
\Re \left(
\frac{\partial \widetilde{a}_{\ell' m'}^{ \left( \beta \right) }}
{\partial a_{\ell''' -m'''}^{ \left( \varepsilon \right) }}
\right) \right] \\
&~~~~~~~~~~~~~~~~~~~~~~~~~~~~~~~ \times
\mathrm{Cov} \left(
\Re \left(
a_{\ell'' m''}^{ \left( \gamma \right) }
\right),
\Re \left(
a_{\ell''' m'''}^{ \left( \varepsilon \right) }
\right) \right) \\
&~~~~~~~~~~~~~~~~~~~~~~~~~~~~~~~ +
\left[ \Im \left(
\frac{\partial \widetilde{a}_{\ell m}^{ \left( \alpha \right) }}
{\partial a_{\ell'' m''}^{ \left( \gamma \right) }}
\right)
- \left( -1 \right)^{m''}
\Im \left(
\frac{\partial \widetilde{a}_{\ell m}^{ \left( \alpha \right) }}
{\partial a_{\ell'' -m''}^{ \left( \gamma \right) }}
\right) \right]
\left[ \Im \left(
\frac{\partial \widetilde{a}_{\ell' m'}^{ \left( \beta \right) }}
{\partial a_{\ell''' m'''}^{ \left( \varepsilon \right) }}
\right)
- \left( -1 \right)^{m'''}
\Im \left(
\frac{\partial \widetilde{a}_{\ell' m'}^{ \left( \beta \right) }}
{\partial a_{\ell''' -m'''}^{ \left( \varepsilon \right) }}
\right) \right] \\
&~~~~~~~~~~~~~~~~~~~~~~~~~~~~~~~ \times
\mathrm{Cov} \left(
\Im \left(
a_{\ell'' m''}^{ \left( \gamma \right) }
\right),
\Im \left(
a_{\ell''' m'''}^{ \left( \varepsilon \right) }
\right) \right) \Bigg) \Bigg].
\label{Eqn:cov_re_re_wfullsky}
\end{split}
\end{equation}
The full-sky covariances are given by
\begin{equation}
\mathrm{Cov} \left(
\Re \left( a_{\ell m}^{ \left( \alpha \right) } \right),
\Re \left( a_{\ell' m'}^{ \left( \beta \right) } \right)
\right)
=
\begin{cases}
C_{\ell}^{\alpha \beta} \delta_{\ell \ell'} \delta_{m m'}
& m = 0; \\
C_{\ell}^{\alpha \beta} \delta_{\ell \ell'} \delta_{m m'} / 2
& m > 0,
\label{Eqn:full_sky_cov_rere}
\end{cases}
\end{equation}
and for $m > 0$,
\begin{equation}
\mathrm{Cov} \left(
\Im \left( a_{\ell m}^{ \left( \alpha \right) } \right),
\Im \left( a_{\ell' m'}^{ \left( \beta \right) } \right)
\right)
= C_{\ell}^{\alpha \beta} \delta_{\ell \ell'} \delta_{m m'} / 2.
\label{Eqn:full_sky_cov_imim}
\end{equation}
Inserting these into Equation \eqref{Eqn:cov_re_re_wfullsky}, evaluating the delta functions and rearranging, we obtain
\begin{equation}
\begin{split}
&
\mathrm{Cov} \left(
\Re \left( \widetilde{a}_{\ell m}^{ \left( \alpha \right) } \right),
\Re \left( \widetilde{a}_{\ell' m'}^{ \left( \beta \right) } \right)
\right) \\
&= \sum_{\gamma, \varepsilon}
\sum_{\ell''}
C_{\ell''}^{\gamma \varepsilon}
\Bigg[
\Re \left(
\frac{\partial \widetilde{a}_{\ell m}^{ \left( \alpha \right) }}
{\partial a_{\ell'' 0}^{ \left( \gamma \right) }}
\right)
\Re \left(
\frac{\partial \widetilde{a}_{\ell' m'}^{ \left( \beta \right) }}
{\partial a_{\ell'' 0}^{ \left( \varepsilon \right) }}
\right)
\\
&~~~~~~~~~~~~~~~~~~~~~~~~~ +
\frac{1}{2}
\sum_{m'' > 0} \Bigg(
\left[
\Re \left(
\frac{\partial \widetilde{a}_{\ell m}^{ \left( \alpha \right) }}
{\partial a_{\ell'' m''}^{ \left( \gamma \right) }}
\right)
+ \left( -1 \right)^{m''}
\Re \left(
\frac{\partial \widetilde{a}_{\ell m}^{ \left( \alpha \right) }}
{\partial a_{\ell'' -m''}^{ \left( \gamma \right) }}
\right) \right]
\left[
\Re \left(
\frac{\partial \widetilde{a}_{\ell' m'}^{ \left( \beta \right) }}
{\partial a_{\ell'' m''}^{ \left( \varepsilon \right) }}
\right)
+ \left( -1 \right)^{m''}
\Re \left(
\frac{\partial \widetilde{a}_{\ell' m'}^{ \left( \beta \right) }}
{\partial a_{\ell'' -m''}^{ \left( \varepsilon \right) }}
\right) \right] \\
&~~~~~~~~~~~~~~~~~~~~~~~~~~~~~~~~~~~~~~~~\, +
\left[ \Im \left(
\frac{\partial \widetilde{a}_{\ell m}^{ \left( \alpha \right) }}
{\partial a_{\ell'' m''}^{ \left( \gamma \right) }}
\right)
- \left( -1 \right)^{m''}
\Im \left(
\frac{\partial \widetilde{a}_{\ell m}^{ \left( \alpha \right) }}
{\partial a_{\ell'' -m''}^{ \left( \gamma \right) }}
\right) \right]
\left[ \Im \left(
\frac{\partial \widetilde{a}_{\ell' m'}^{ \left( \beta \right) }}
{\partial a_{\ell'' m''}^{ \left( \varepsilon \right) }}
\right)
- \left( -1 \right)^{m''}
\Im \left(
\frac{\partial \widetilde{a}_{\ell' m'}^{ \left( \beta \right) }}
{\partial a_{\ell'' -m''}^{ \left( \varepsilon \right) }}
\right) \right]
\Bigg) \Bigg].
\end{split}
\end{equation}
Finally, we use the identities that for complex $A$ and $B$,
\begin{equation}
\Re A \Re B + \Im A \Im B = \Re \left( A^* B \right); \quad \quad
\Re A \Re B - \Im A \Im B = \Re \left( A B \right)
\end{equation}
to obtain the final covariance given in Equation \eqref{Eqn:cov_re_re_general}. The variance of a single real part is a special case of Equation \eqref{Eqn:cov_re_re_general} having $\alpha = \beta$, $\ell = \ell'$ and $m = m'$.

\section{Equivalence of the two forms of the joint characteristic function}
\label{App:cf_equivalance}

Here we demonstrate that the expression for the joint characteristic function used in our implementation of the likelihood, given in Equation \eqref{Eqn:CF_alt} and denoted here by $\varphi_\mathrm{alt}$, is mathematically equivalent to the known analytic form from \citet{Good1963}, given in Equation \eqref{Eqn:joint_cf} and denoted here by $\varphi_\mathrm{Good}$.

The form given in \citet{Good1963} is written in terms of the determinant of the complex square matrix which we will denote by $\mathbfss{X}$:
\begin{equation}
\mathbfss{X} =
\mathbfss{I} - 2i \sum_\ell \sum_{\alpha \beta}
t_\ell^{\alpha \beta} \mathbfss{M}_\ell^{\alpha \beta} \bm{\Sigma},
\end{equation}
where $\mathbfss{I}$ is the identity matrix. The determinant of any matrix is equal to the product of its eigenvalues, so we may write $\varphi_\mathrm{Good}$ as
\begin{equation}
\varphi_\mathrm{Good} \left( \mathbfit{t} \right) =
\prod_j \lambda_j^{-1/2}, \quad \quad
\lambda_j \in \lambda \left( \mathbfss{X} \right).
\end{equation}
The eigenvalues of $\mathbfss{X}$, $\lambda \left( \mathbfss{X} \right)$, are defined by the eigenvalue equation,
\begin{equation}
    \lvert \mathbfss{X} - \lambda_j \mathbfss{I} \rvert = 0 \quad
    \forall ~ \lambda_j \in \lambda \left( \mathbfss{X} \right).
    \label{Eqn:x_evs}
\end{equation}
Now let us define another matrix $\mathbfss{Y}$ such that $\mathbfss{X} = \mathbfss{I} - \mathbfss{Y}$. Inserting this into Equation \eqref{Eqn:x_evs}, we obtain
\begin{equation}
    \Big\lvert
    - \left[ \mathbfss{Y} - \left( 1 - \lambda_j \right) \mathbfss{I}\right]
    \Big\rvert = 0 \quad
    \forall ~ \lambda_j \in \lambda \left( \mathbfss{X} \right).
    \label{Eqn:det_y}
\end{equation}
For any matrix $\mathbfss{A}$, the determinant of $- \mathbfss{A}$ is given by
\begin{equation}
    \lvert - \mathbfss{A} \rvert = \pm \lvert \mathbfss{A} \rvert,
\end{equation}
depending on whether the rank of $\mathbfss{A}$ is even or odd. Using this fact in Equation \eqref{Eqn:det_y} implies that
\begin{equation}
    \Big\lvert
    \mathbfss{Y} - \left( 1 - \lambda_j \right) \mathbfss{I}
    \Big\rvert = 0 \quad
    \forall ~ \lambda_j \in \lambda \left( \mathbfss{X} \right).
\end{equation}
Therefore, the eigenvalues of $\mathbfss{X}$ and $\mathbfss{Y}$ are related as
\begin{equation}
\lambda \left( \mathbfss{Y} \right) = 1 - \lambda \left( \mathbfss{X} \right),
\quad \quad \text{and hence} \quad \quad
\lambda \left( \mathbfss{X} \right) = 1 - \lambda \left( \mathbfss{Y} \right).
\end{equation}
This allows us to write $\varphi_\mathrm{Good}$ in terms of the eigenvalues of $\mathbfss{Y}$,
\begin{equation}
\varphi_\mathrm{Good} \left( \mathbfit{t} \right) =
\prod_j \left( 1 - \lambda_j \right)^{-1/2}, \quad \quad
\lambda_j \in \lambda \left( \mathbfss{Y} \right).
\label{Eqn:Good_evs_y}
\end{equation}
Finally, we use the fact that the eigenvalues of a scalar multiple of a matrix are equal to the scalar multiplied by the original matrix,
\begin{equation}
    \lambda \left( \alpha \mathbfss{A} \right) =
    \alpha \lambda \left( \mathbfss{A} \right),
\end{equation}
to extract the multiple of $2i$ from $\mathbfss{Y}$:
\begin{equation}
    \lambda \left( \mathbfss{Y} \right) =
    \lambda \left(
    2i \sum_\ell \sum_{\alpha \beta}
    t_\ell^{\alpha \beta} \mathbfss{M}_\ell^{\alpha \beta} \bm{\Sigma}
    \right)
    = 2i \times \lambda \left(
    \sum_\ell \sum_{\alpha \beta}
    t_\ell^{\alpha \beta} \mathbfss{M}_\ell^{\alpha \beta} \bm{\Sigma}
    \right),
\end{equation}
which we insert into Equation \eqref{Eqn:Good_evs_y} to obtain our alternative form,
\begin{equation}
\begin{split}
    \varphi_\mathrm{Good} \left( \mathbfit{t} \right) &=
    \prod_j \left( 1 - 2i \lambda_j \right)^{-1/2}, \quad \quad
    \lambda_j \in \lambda \left( \sum_\ell \sum_{\alpha \beta}
    t_\ell^{\alpha \beta} \mathbfss{M}_\ell^{\alpha \beta} \bm{\Sigma} \right) \\
    &= \varphi_\mathrm{alt} \left( \mathbfit{t} \right).
\end{split}
\end{equation}


\bsp	
\label{lastpage}
\end{document}